\documentclass[a4paper,11pt]{article}
\pdfoutput=1 

\usepackage{jcappub} 

\usepackage[T1]{fontenc} 

\usepackage{empheq}

\usepackage{listings}

\usepackage{tablefootnote}

\title{\boldmath Analytical Template for the 4-Point Correlation Function Covariance Beyond the Gaussian Random Field ${\rm II}$: 1-Loop Corrections with Third-Order Densities}
\author[1]{William Ortolá Leonard} 
\author[2]{and Zachary Slepian}

\affiliation[1]{Department of Physics, University of Florida,\\2001 Museum Rd., Gainesville, FL 32611, USA}
\affiliation[2]{Department of Astronomy, University of Florida,\\211 Bryant Space Science Center, Gainesville, FL 32611, USA}

\emailAdd{wortola@ufl.edu}
\emailAdd{zslepian@ufl.edu}

\abstract{Analytical templates for the 4-Point Correlation Function (4PCF) covariance matrix have been developed in the past assuming a Gaussian Random Field (GRF). In this work, we present the second part of the beyond GRF calculation of the 4PCF covariance, incorporating 1-loop corrections stemming from the third-order density contrast. Furthermore, we introduce a non-trivial galaxy biasing scheme at third order. To simplify the calculation, we decompose the covariance into three distinct structures and leverage the isotropic basis of \cite{Cahn_Iso}. This approach reduces the complexity of the high-dimensional integrals that would be naively involved, enabling the angular parts to be performed and leaving us with low-dimensional radial integrals. This analytical template could provide a more accurate characterization of the statistical errors on the 4PCF in future surveys, improving our ability to probe both its parity-even and parity-odd modes. This is the second and final paper in a two-part series.}

\usepackage{empheq}
\usepackage{listings}
\usepackage{hyperref}
\usepackage{tablefootnote}
\usepackage{graphicx} 

\newcommand{\vecx}{\mathbf{x}}
\newcommand{\vecr}{\mathbf{r}}
\newcommand{\vecs}{\mathbf{s}}
\newcommand{\veck}{\mathbf{k}}
\newcommand{\vecq}{\mathbf{q}}
\newcommand{\dg}{\delta_{\rm g}}
\newcommand{\dD}{\delta_{\rm D}^{[3]}}
\newcommand{\dK}{\delta^{\rm K}}
\newcommand{\dlin}{\delta_{\rm lin.}}
\newcommand*\widefbox[1]{\fbox{\hspace{1em}#1\hspace{1em}}}
\newcommand{\hatk}{\widehat{\mathbf{k}}}
\newcommand{\hats}{\widehat{\mathbf{s}}}
\newcommand{\hatr}{\widehat{\mathbf{r}}}
\newcommand{\hatq}{\widehat{\mathbf{q}}}
\newcommand{\Pk}{P_{\rm lin}}
\newcommand{\PP}{\mathcal{P}}
\newcommand{\W}{W^{(3)}}

\newcommand{\Cth}{{\rm Cov}_{4,\text{1L}}^{[3]}(\mathbf{R},\mathbf{R}')}

\newcommand{\hatR}{\mathbf{\widehat{R}}}

\newcommand{\tj}[6]{\begin{pmatrix} {#1} & {#2} & {#3}\\ {#4} & {#5} & {#6}\end{pmatrix}}
\begin{document}
\maketitle
\flushbottom

\section{Introduction}
The cosmic web encodes rich higher-order information about gravitational dynamics and potential departures from fundamental symmetries. Recent 4-Point Correlation Function (4PCF) measurements using data from the Sloan Digital Sky Survey (SDSS) Baryon Oscillation Spectroscopic Survey (BOSS) and the Dark Energy Spectroscopic Instrument (DESI) have reported hints of parity-odd structure \cite{Hou_Parity, Philcox_Parity, Slepian_Parity_odd}, using estimators based on \cite{cahn_prl} and the isotropic basis of \cite{Cahn_Iso}. In interpreting these measurements, previous analyses have made use of the Gaussian theoretical 4PCF covariance model \cite{Hou_Cov}, as well as hybrid covariance estimates, in which the analytical model is combined with mocks to compress the information content and to incorporate some nonlinear effects. The even-parity 4PCF has also been measured in BOSS \cite{Philcox_4PCF_measurement} and DESI \cite{Hou_4PCF_Measurement}. To robustly assess the cosmological implications of the 4PCF, a more complete theoretical description of the covariance beyond the Gaussian limit may therefore be required.

DESI Y1 \cite{DESI_DR1} dramatically increases the statistical power relative to BOSS, sharpening both the promise and the demands on covariance modeling. Building on advances in beyond-Gaussian covariance for two-point statistics \cite{Bertolini_Cov, Mohammed_Cov, Wadekar_Cov}, here we present the 1-loop (next-to-leading-order) 4PCF covariance contributions involving third-order density fields in SPT, together with a third-order galaxy-bias treatment. These terms are complementary to the second-order density and galaxy-bias contributions computed in \cite{Ortola_4PCF_CovI}; the complete NLO 4PCF covariance model requires both sets of contributions to be included. Together, they extend the fidelity of the covariance modeling into the mildly nonlinear regime and capture mode-coupling effects that are neglected in the Gaussian Random Field (GRF) approximation. The goal of this series of papers is to enable a more accurate propagation of uncertainties for 4PCF-based parity tests and cosmological inference. A detailed assessment of the impact of these NLO contributions on parameter inference and detection significances is left for future work.

The structure of this paper is as follows. In $\S$\ref{sec:Cov_Compu}, we set up the next-to-leading-order covariance matrix, at tenth order in the density field with third-order densities. In $\S$\ref{sec:Resulting_Cov}, we show how to mathematically model the covariance we previously set in $\S$\ref{sec:Cov_Compu}. Then, we conclude by summarizing our results and discussing future work in $\S$\ref{sec:Conclussion}. Finally, we use the appendices to develop all mathematical tools needed on $\S$\ref{sec:Resulting_Cov}. 

\section{Setting Up the 1-Loop Covariance with Third-Order Densities} \label{sec:Cov_Compu}
The covariance is defined as \cite{Hou_Cov}:
\begin{align}
{\rm Cov}(\zeta(\mathbf{R}),\zeta(\mathbf{R}')) \equiv \left< \zeta(\mathbf{R})\zeta(\mathbf{R}')\right> - \big<\zeta(\mathbf{R})\big> \left<\zeta(\mathbf{R}')\right>,
\end{align}
where $\zeta$ represents any N-Point Correlation Function (NPCF) we wish to evaluate and the angle bracket represents the ensemble average over realizations of the cosmic density field. We define $\mathbf{R} \equiv (\vecr_0,\vecr_1,\cdots,\vecr_{N-1})$, and $\mathbf{R}'$ in the same way. Following the procedure of \cite{Hou_Cov}, we evaluate the first term:
\begin{align}\label{eq:Con_Cov_First_Expansion}
\left< \zeta(\mathbf{R})\zeta(\mathbf{R}')\right> = \int\frac{ d^3\vecs}{V} \left< \prod_{i=0}^{N-1} \;\dg(\vecx+\vecr_i)\;\dg(\vecx+\vecr'_i+ \vecs) \right>,
\end{align}
where the angle brackets represent the ensemble average of $\vecx$. We set $N=4$, \textit{i.e.,} we evaluate the 4PCF covariance. Subscript ``g'' indicates the density fluctuations, $\delta$, are for galaxies, and $V$ is the survey volume. Using the Eulerian bias expansion \cite{Bernardeau, Scoccimarro, Desjacques_bias_review}, we have:
\begin{align}\label{eq:dc_galaxy_expansion}
&\dg(\vecx) = b_0 + b_1\delta_{\rm m}(\vecx) + \frac{b_2}{2}\delta_{\rm m}^{2}(\vecx) \ + b_{S} \;S^{(2)}(\vecx)\nonumber \\
&\qquad \;\;\;\;+ \frac{b_3}{6} \delta_{\rm m}^{3}(\vecx)+  b_{\delta\mathcal{G}_2} \delta(\vecx) \mathcal{G}^{(2)}(\vecx) + b_{\mathcal{G}_3}\mathcal{G}^{(3)}(\vecx) +b_{\Gamma_3}\Gamma^{(3)}(\vecx),
\end{align}
with $b_0$ ensuring $\langle\dg\rangle = 0$, but will be omitted in the rest of this work since it does not enter connected correlation functions \cite{Scoccimarro}. The subscript ``m'' indicates the matter density fluctuation. Then, the matter density contrast is expanded using SPT:
\begin{align}\label{eq:dc_matter_expansion}
\delta_{\rm m}(\vecx) = \delta_{\rm lin.} (\vecx)+ \delta^{(2)}(\vecx) + \delta^{(3)}(\vecx) + \mathcal{O}(\delta_{\rm lin.}^4).
\end{align}
Above, we expand the matter density contrast up to third order in the linear density contrast, $\delta_{\rm lin.}$; we define $\delta^{(3)}(\vecx)$, $\delta^{3}_{\rm lin.}(\vecx)$, $\mathcal{G}^{(2)}(\vecx)$, $\mathcal{G}^{(3)}(\vecx)$ and $\Gamma^{(3)}(\vecx)$ via their inverse Fourier Transforms (FTs) in Appendix \ref{sec:Generalization_delta_3}. In \cite{Ortola_4PCF_CovI}, we have already evaluated the contributions entering at second order; the first line of Eq. (\ref{eq:dc_galaxy_expansion}). In this work we focus on the contributions entering at third order; the second line of Eq. (\ref{eq:dc_galaxy_expansion}). Hence, inserting Eq. (\ref{eq:dc_matter_expansion}) into Eq. (\ref{eq:dc_galaxy_expansion}), and incorporating the resultant expression into Eq. (\ref{eq:Con_Cov_First_Expansion}) for $N=4$, we obtain:
\begin{align}\label{eq:Con_Cov_second_Expansion}
&\left< \zeta(\mathbf{R})\zeta(\mathbf{R}')\right> = \int\frac{ d^3\vecs}{V} \Bigg\langle
\prod_{i=0}^{3}
\bigg[ b_1\left(\delta_{\rm lin.} + \delta^{(2)} + \delta^{(3)}\right) + \frac{b_3}{6}\left(\delta_{\rm lin.} + \delta^{(2)} + \delta^{(3)}\right)^3  \nonumber \\
&\qquad \qquad \qquad  + b_{\delta\mathcal{G}_2} \left(\delta_{\rm lin.} + \delta^{(2)} + \delta^{(3)}\right)\mathcal{G}^{(2)} + b_{\mathcal{G}_3}\mathcal{G}^{(3)} +b_{\Gamma_3}\Gamma^{(3)} \bigg](\mathbf{x}+\mathbf{r}_i)\nonumber \\
&\qquad\qquad\qquad\times
\bigg[ b_1\left(\delta_{\rm lin.} + \delta^{(2)} + \delta^{(3)}\right) + \frac{b_3}{6}\left(\delta_{\rm lin.} + \delta^{(2)} + \delta^{(3)}\right)^3  \nonumber \\
&\qquad \qquad \qquad  + b_{\delta\mathcal{G}_2} \left(\delta_{\rm lin.} + \delta^{(2)} + \delta^{(3)}\right)\mathcal{G}^{(2)}  + b_{\mathcal{G}_3}\mathcal{G}^{(3)} +b_{\Gamma_3}\Gamma^{(3)}\bigg](\mathbf{x}+\mathbf{r}'_{i}+\mathbf{s})\Bigg\rangle.
\end{align}
Here, the arguments $\vecx+\vecr_i$ and $\vecx+\vecr'_i+\vecs$ apply to all terms inside the square brackets. Carrying out the product, we obtain the 1-loop covariance with third-order densities as:
\begin{align}\label{eq:Full_cov}
&\Cth=\nonumber \\ 
& \qquad\quad b_1^8  \bigg\{ 
 \sum_{i=0}^{3} \bigg< 
   \delta^{(3)}(\vecx + \vecr_{i})
   \prod_{p=0,p\neq i}^3\delta_{\rm lin.}(\vecx+\vecr_p)
   \prod_{t=0}^{3} \delta_{\rm lin.}(\vecx+\vecr'_t+ \vecs)
 \bigg> + {\rm symm.} 
\bigg\} \nonumber \\ 
& \qquad + b_1^7\frac{b_3}{6}  \bigg\{ 
 \sum_{i=0}^{3} \bigg< 
   \delta^{3}_{\rm lin.}(\vecx + \vecr_{i})
   \prod_{p=0,p\neq i}^3\delta_{\rm lin.}(\vecx+\vecr_p)
   \prod_{t=0}^{3} \delta_{\rm lin.}(\vecx+\vecr'_t+ \vecs)
 \bigg> + {\rm symm.} 
\bigg\} \nonumber \\ 
& \qquad + b_1^7 b_{\delta\mathcal{G}_2} \bigg\{ 
 \sum_{i=0}^{3} \bigg< 
   \delta_{\rm lin.}(\vecx + \vecr_{i})\,
   \mathcal{G}^{(2)}(\vecx + \vecr_{i})
   \prod_{p=0,p\neq i}^3 \delta_{\rm lin.}(\vecx+\vecr_p)
   \prod_{t=0}^{3} \delta_{\rm lin.}(\vecx+\vecr'_t+ \vecs)
 \bigg> + {\rm symm.} 
\bigg\} \nonumber \\ 
& \qquad + b_1^7 b_{\mathcal{G}_3}  \bigg\{ 
 \sum_{i=0}^{3} \bigg< 
   \mathcal{G}^{(3)}(\vecx + \vecr_{i})
   \prod_{p=0,p\neq i}^3\delta_{\rm lin.}(\vecx+\vecr_p)
   \prod_{t=0}^{3} \delta_{\rm lin.}(\vecx+\vecr'_t+ \vecs)
 \bigg> + {\rm symm.} 
\bigg\}\nonumber \\  
& \qquad + b_1^7 b_{\Gamma_3}  \bigg\{ 
 \sum_{i=0}^{3} \bigg< 
   \Gamma^{(3)}(\vecx + \vecr_{i})
   \prod_{p\neq i}\delta_{\rm lin.}(\vecx+\vecr_p)
   \prod_{t=0}^{3} \delta_{\rm lin.}(\vecx+\vecr'_t+ \vecs)
 \bigg> + {\rm symm.} 
\bigg\}. 
\end{align}
Above, we used the subscript 4,1L on the left-hand side to denote our evaluation of the 4PCF 1-loop covariance and the superscript $[3]$ to indicate that we are only evaluating contributions from the third-order density field. This equation captures all possible permutations of the terms through summation and two products, ensuring that for each choice of $i$, $p \in \{0,1,2,3\}$ must be distinct. In what follows, we display only the product over $p$, with the understanding that $p \neq i$ is implicitly enforced. To restore the symmetry between the unprimed and primed vectors, an additional term denoted ``+ symm.'' has been included; we show an example of how to compute these terms below: 
\begin{align}
& b_1^8  \bigg\{ 
 \sum_{i=0}^{3} \bigg< 
   \delta^{(3)}(\vecx + \vecr_{i})
   \prod_{p=0,p\neq i}^3\delta_{\rm lin.}(\vecx+\vecr_p)
   \prod_{t=0}^{3} \delta_{\rm lin.}(\vecx+\vecr'_t+ \vecs)
 \bigg> + {\rm symm.} \bigg\} \nonumber \\
& \quad = b_1^8  \bigg\{ 
 \sum_{i=0}^{3} \bigg< 
   \delta^{(3)}(\vecx + \vecr_{i})
   \prod_{p=0,p\neq i}^3\delta_{\rm lin.}(\vecx+\vecr_p)
   \prod_{t=0}^{3} \delta_{\rm lin.}(\vecx+\vecr'_t+ \vecs)
 \bigg>\nonumber \\
& \quad + \sum_{i=0}^{3} \bigg< 
   \delta^{(3)}(\vecx + \vecr'_{i}+\vecs)
   \prod_{p=0,p\neq i}^3\delta_{\rm lin.}(\vecx+\vecr'_p+\vecs)
   \prod_{t=0}^{3} \delta_{\rm lin.}(\vecx+\vecr_t)
 \bigg> \bigg\}.
\end{align}
We prove in Appendix \ref{sec:Generalization_delta_3} that the five terms in Eq. (\ref{eq:Full_cov}) can all be written in terms of a simple expression, using a generalization of the Fourier space, $F^{(3)}$ kernel, that we develop there. 

Using the result of Appendix \ref{sec:Generalization_delta_3} and rewriting the density contrast via its inverse FT, the 1-loop covariance is:
\begin{align}
&\Cth = \prod_{v=0}^{3}\int_{\vecs}\int_{\;\veck_v}\int_{ \; \veck'_v}\left[ \;e^{-i\veck_v\cdot(\vecx+\vecr_v)} e^{-i\veck'_v \cdot(\vecx+\vecr'_v+\vecs)} \right]\nonumber \\
& \qquad \qquad \qquad \times \sum_{\mu} \sum_{i}\left< \widetilde{\delta}_\mu^{(3)}(\veck_i)\prod_{p,t}\widetilde{\delta}_{\rm lin.}(\veck_p) \widetilde{\delta}_{\rm lin.}(\veck'_t)\right>,
\end{align}
where $\mu$ runs over $\{0,1,2,3,4\}$ such that $ \left\{\widetilde{\delta}_0^{(3)},\widetilde{\delta}_1^{(3)},\widetilde{\delta}_2^{(3)},\widetilde{\delta}_3^{(3)},\widetilde{\delta}_4^{(3)}\right\} = \left\{\delta^{(3)},\delta_{\rm lin.}^{3}, \delta\mathcal{G}^{(2)},\mathcal{G}^{(3)}, \Gamma^{(3)} \right\}$, and as already mentioned below Eq. (\ref{eq:Full_cov}) $i\neq p$. We have abbreviated the 3D integrals as:
\begin{align}
&\int_{\vecs} = \int \frac{d^{3}\vecs}{V}\;\; {\rm and} \;\;\int_{\veck} = \int \frac{d^{3}\veck}{(2\pi)^3}.\nonumber 
\end{align}

\begin{figure}[h!]
\centering
\includegraphics[scale=0.25]{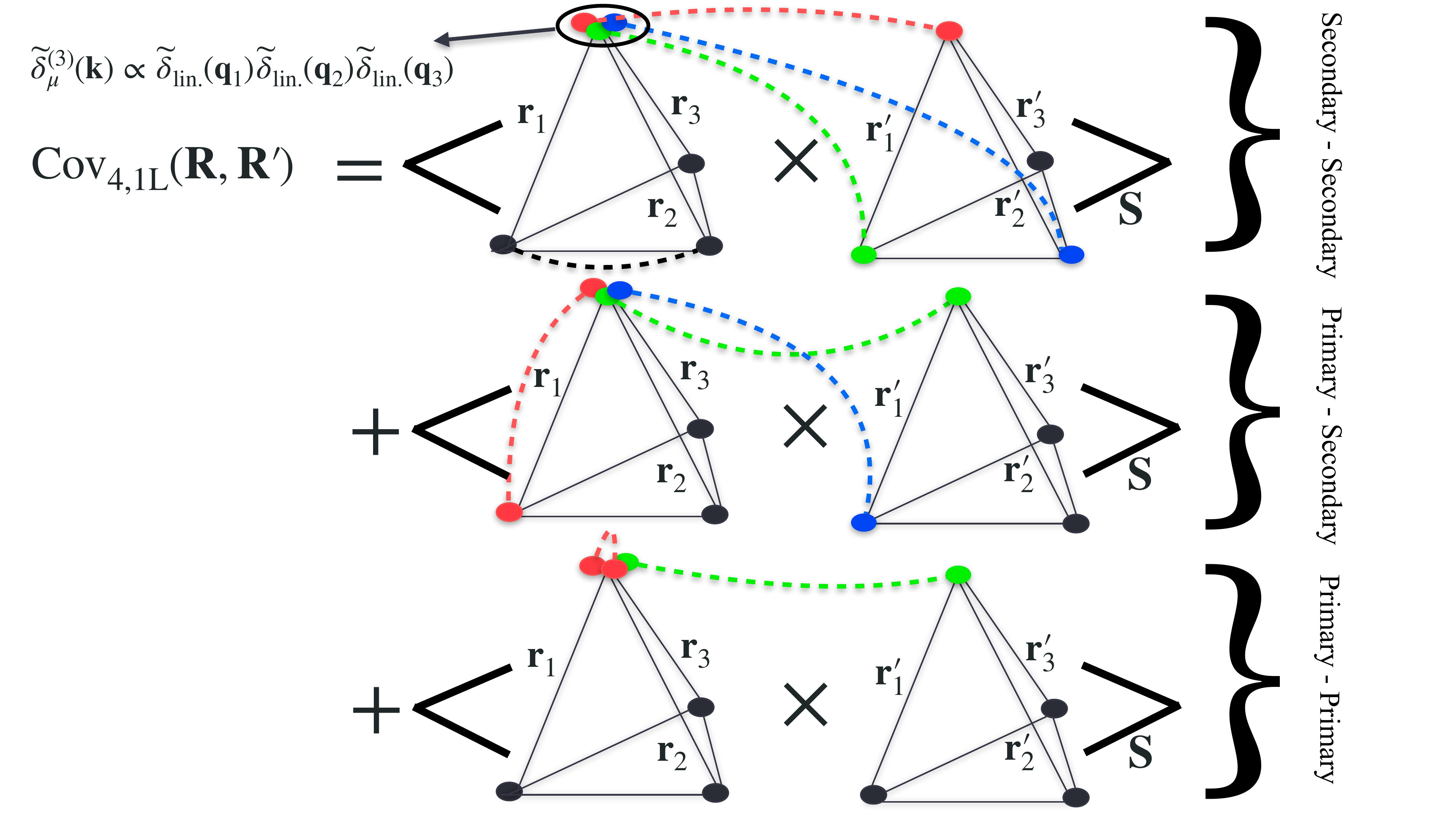}
\caption{Illustration of the three different possible configurations for the covariance matrix at 1-loop order with third-order densities. The three  configurations are: Secondary-Secondary, Primary-Secondary and Primary-Primary. The dashed lines represent the contraction of two density contrasts via Wick's theorem. We name each configuration based on the how the two density contrasts within the same tetrahedron are contracted together. We define the ‘primary’ points as those that contain the third-order densities and depict them in the above diagram with three colored dots. We define the ‘secondary’ points as those that contain the linear densities and depict them with one black dot. The above diagram does not show Wick contractions between the leftover secondary densities (black dots), but what results from them has been explicitly included and calculated in each of the sections below. However, in the first structure we do explicitly show the Wick contraction between two of the secondary densities since they get contracted within the same tetrahedron. The primary densities are $\vecq_1,\;\vecq_2$, and $\vecq_3$ from left to right in the unprimed tetrahedron. In the main text, we introduce $\vecr_0 = \vecr'_0 = 0$ to restore the symmetry between arguments of the calculations in $\S$\ref{sec:Resulting_Cov}.}
\label{fig:dc_configurations}
\end{figure}

Expanding the third-order density field using Eq. (\ref{eq:delta_i_def}), we obtain:
\begin{align}\label{eq:sub_cov_before_Wicks}
&\Cth = \prod_{v=0}^{3}\int_{\vecs}\int_{\;\veck_v}\int_{ \; \veck'_v}\left[ \;e^{-i\veck_v\cdot(\vecx+\vecr_v)} e^{-i\veck'_v \cdot(\vecx+\vecr'_v+\vecs)} \right]\nonumber \\
& \qquad \qquad  \times \sum_{\mu} \int d^3\vecq_1 \int d^3\vecq_2\; d^3\vecq_3  \;W_{\mu}^{(3)}(\vecq_1,\vecq_2,\vecq_3) \;\dD(\veck_0-\vecq_1-\vecq_2-\vecq_3)\; \nonumber \\
& \qquad \qquad  \times  \left< \widetilde{\delta}_{\rm lin.}(\vecq_1)\widetilde{\delta}_{\rm lin.}(\vecq_2)\widetilde{\delta}_{\rm lin.}(\vecq_3)\prod_{p\neq0,t}\widetilde{\delta}_{\rm lin.}(\veck_p) \widetilde{\delta}_{\rm lin.}(\veck'_t)\right>,
\end{align}
where the $W^{(3)}$ kernel is defined in Eq. (\ref{eq:W_3_def}). We may now evaluate the ensemble average of the density contrasts using Wick's theorem in three different cases, as shown in Fig. \ref{fig:dc_configurations}. In the next section, we pursue this to develop the basic mathematical elements of the covariance.

\begin{table} [h!]
\centering
\begin{tabular}{ |p{2.5cm}|p{1.5cm}| p{10cm}|}
 \hline
 \multicolumn{3}{|c|}{Table of Coefficients} \\
 \hline
 Coefficient & Eq./Sec. & Definition \\
 \hline
 &&\\
 $C_{L_1,L_2,L_3}$ & \ref{eq:Constant_from_exp} & Coefficient of plane-wave expansion when projected onto the isotropic basis functions. \\
  &&\\
 $\Upsilon_{L_1,L_2,L_3}$ & \ref{eq:Upsilon_def} & Coefficient from the splitting of position-space-vector and wave-vector isotropic basis functions. \\ 
  &&\\
  $s^{(\rm I)}_{\ell}$ & \ref{eq:def_I_k1} & Coefficient from expressing two spherical harmonics as an isotropic basis function. \\ 
&&\\
   $c^{(\mu)}_{j,n}$  &  \ref{eq:W2_def} & $W_{\mu}^{(2)}$ kernel coefficients.  \\
  &&\\
  $D_j$ & \ref{eq:D_j_def} & Coefficient from expressing an isotropic basis function as two spherical harmonics. \\
 &&\\
 $\mathcal{D}_{n'_1,n'_2,n'_3,n'_4}^{n_{12},n_{13},n_{23}}$ & \ref{eq:W_3_def} & Coefficient from the Generalized $W^{(3)}$ kernel, ensuring only the correct terms of the third-order kernel are non-zero. \\
 &&\\
 $d_j^{(n)}$ & \ref{eq:exp_sph} & Coefficient from expressing a dot porduct in terms of the isotropic basis function. \\
 &&\\
 $ \mathcal{G}_{L_1,L_2,L_3} $ &   \ref{eq:modified_Gaunt} & Modified Gaunt integral from evaluation of a 3-argument isotropic basis function. \\ 
  &&\\
  $ \overline{\mathcal{H}}_{L_1,L_2,M_1,M_2}^{\ell_1,\ell_2,m_1,m_2}$& \ref{eq:Def_H_bar}& Coefficient from the integration of four spherical harmonic. \\ 
  &&\\
  $ \mathcal{J}_{L_1,L_2,L_3}^{\ell_1,\ell_2,m_1,m_2}$&\ref{eq:Fancy_J_def}&Coefficient from the integration of one 3-argument isotropic basis function times two spherical harmonics.\\ 
  &&\\
  $\mathcal{L}_{L_1,L_2,L_3}^{j_1,j_2,j_3,j_4,\{m_{j_i}\}}\;$ &\ref{eq:Fancy_L_def} & Coefficient from the integration of a 3-argument isotropic basis functions times four spherical harmonics. \\
  &&\\
  $\mathcal{\overline{Q}}^{\Lambda,\Lambda_s,\Lambda'}$ &\ref{Sec:Angular_Integrals} & Coefficients from the evaluation of the $\hats$ integrals for each of the configurations shown in Fig. \ref{fig:dc_configurations}. \\
   &&\\
 \hline
\end{tabular} 
\caption{Table with the coefficients relevant for the computation of the covariance.}
\label{table:1}
\end{table}

\section{Modeling of the 1-Loop Covariance with Third-Order Densities}\label{sec:Resulting_Cov}
In this section, we derive the 1-loop covariance using the third-order density contrast, for which all possible configurations have been depicted in Figure \ref{fig:dc_configurations}. We show that only these terms enter the 1-loop covariance in Appendix \ref{sec:Forming_conn_4PCF}. We classify these configurations according to how the loop, involving the third-order density, connects the vertices of the tetrahedron. If the loop joins two densities that both come from the linear fields, we call it secondary-secondary; if it connects a third-order (primary) density to a linear (secondary) one, we call it primary-secondary; and if it links two third-order densities, it becomes primary-primary. This distinction simply reflects which type of density participates in closing the loop. We begin the analysis with the secondary-secondary configuration shown in Figure \ref{fig:dc_configurations}.

\subsection{Secondary-Secondary Configurations}\label{sec:S-S_conf}

Using Wick's theorem for the secondary-secondary configuration shown in Fig. \ref{fig:dc_configurations} on Eq. (\ref{eq:sub_cov_before_Wicks}) results in:
\begin{align}
&{\rm Cov}_{4, \rm 1L}^{[3],{\rm S-S}}(\mathbf{R},\mathbf{R}') = \prod_{v=0}^{3} \int_{\vecs}\int_{\veck_v}\int_{\veck'_v}\left[ \;e^{-i\veck_v\cdot(\vecx+\vecr_v)} e^{-i\veck'_v \cdot(\vecx+\vecr'_v+\vecs)} \right]\nonumber \\
& \qquad \qquad \qquad \times\int d^3\vecq_1 \int d^3\vecq_2 \int d^3\vecq_3 \ \;\W(\vecq_1,\vecq_2,\vecq_3) \; \dD(\veck_0-\vecq_1-\vecq_2-\vecq_3)\nonumber \\
& \qquad \qquad \qquad \times  \left<\widetilde{\delta}_{\rm lin.}(\vecq_1)\widetilde{\delta}_{\rm lin.}(\veck'_0)\right>\left<\widetilde{\delta}_{\rm lin.}(\vecq_2)\widetilde{\delta}_{\rm lin.}(\veck'_1)\right> \left<\widetilde{\delta}_{\rm lin.}(\vecq_3)\widetilde{\delta}_{\rm lin.}(\veck'_2)\right>\nonumber \\
& \qquad \qquad \qquad \times\left<\widetilde{\delta}_{\rm lin.}(\veck_3)\widetilde{\delta}_{\rm lin.}(\veck'_3)\right> \left<\widetilde{\delta}_{\rm lin.}(\veck_1)\widetilde{\delta}_{\rm lin.}(\veck_2)\right> + 71\; {\rm perms.},
\end{align}
where $71$ perms. indicates the other 71 combinations\footnote{Finding the number of permutations is done in three steps. i) Start with one of the third-order densities, then pair this third-order density with one of the linear densities from the other tetrahedron; one has 4 possibilities. ii) Repeat with the remaining two third-order densities; one has 6 possibilities. iii) Choose the remaining linear density in the tetrahedron with no third-order densities and pair with one of the three remaining linear densities in the other tetrahedron; one has 3 possibilities. Total number of permutations = 4 $\times$ 6$\times$ 3  = 72 perms.} of Wick's theorem that we can obtain for this secondary-secondary structure; for brevity, we leave this out, but it is implicitly included into our calculations.

Evaluating the ensemble average as power spectra and performing the primed-wave-vectors, the $\veck_0$, and the $\veck_2$ integrals result in:
\begin{align}\label{eq:middlestep_SS_config}
&{\rm Cov}_{4, \rm 1L}^{[3],{\rm S-S}}(\mathbf{R},\mathbf{R}') = (2\pi)^{15} \int_{\rm All} 
e^{-i \vecq_1 \cdot (\vecr_0 - \vecr_0' - \vecs)}
e^{-i \vecq_2 \cdot (\vecr_0 - \vecr_1' - \vecs)}
e^{-i \vecq_3 \cdot (\vecr_0 - \vecr_2' - \vecs)}
e^{-i \veck_{1} \cdot (\vecr_1 - \vecr_2)} \nonumber\\
& \qquad \qquad \times \, e^{-i \veck_3 \cdot (\vecr_3 - \vecr_3' - \vecs)}
\Pk(k_1)\,\Pk(k_3) \Pk(q_1)\, \Pk(q_2)\, \Pk(q_3)\, \,
\W(\vecq_1, \vecq_2, \vecq_3).
\end{align}

As shown in Eq. (\ref{eq:W_3_def}), the third-order kernel has the coupled wave-vector magnitude $q_{23}$ in the denominator; we make use of Eq. (C.7) in \cite{Ortola_4PCF} to decouple it:
\begin{align}\label{eq:Decoupling_two_vects}
\frac{1}{k_{23}^{n_{23}}} = \Omega_{n_{23}} \sum_{\ell_2=0}^{\infty}\sum_{m_2 = -\ell_2}^{\ell_2}(-1)^{\ell_2} Y_{\ell_2,m_2}(\mathbf{\widehat{k}}_2) Y_{\ell_2,m_2}(\hatk_3)\int_{0}^{\infty}dr\; j_{\ell_2}(k_2r)j_{\ell_2}(k_3r) r^{n_{23}-1},
\end{align}
where we have written the angular part in terms of the spherical harmonics and have defined:
\begin{align}
\Omega_{n_{23}}\equiv\frac{8\pi \;i^{n_{23}}}{ \Gamma(n_{23}-1) \left[ (-1)^{n_{23}-1} - 1 \right]}. 
\end{align}
Expanding the third-order kernel with Eq. (\ref{eq:W_3_def}) and using Eq. (\ref{eq:Decoupling_two_vects}),  we can simplify Eq. (\ref{eq:middlestep_SS_config}) by reducing it to the lowest-dimensional integral possible. We start with the $\vecq_2$ integral:
\begin{align}\label{eq:Def_of_I_2}
&I^{(q_2)}(\mathbf{r}_0,-\vecr'_1,-\vecs) \equiv \int_{\vecq_2} q_2^{n'_2+n}\,Y_{j_{12},m_{j_{12}}}(\hatq_2)Y_{j_{23},m_{j_{23}}}(\hatq_2)Y_{j,m_{j}}(\hatq_2) Y_{\ell_2,m_2}(\hatq_2) \nonumber\\
& \qquad \qquad \qquad \qquad \times P\left(q_2\right) \; j_{\ell_2}(q_2r)\; e^{-i \vecq_2 \cdot\left(\mathbf{r}_0-\mathbf{s}-\vecr'_1\right)},
\end{align}
where we have intentionally left out the $j_{12},\;j_{23},\;j$,\; $n$ sums and factor of $4\pi$ since they will also affect the result of the $\vecq_1$ and $\vecq_3$ integrals. We reintroduce these sums in Eq. (\ref{eq:SS_res}). 

\begin{figure}[h!]
\centering
\includegraphics[scale=0.6]{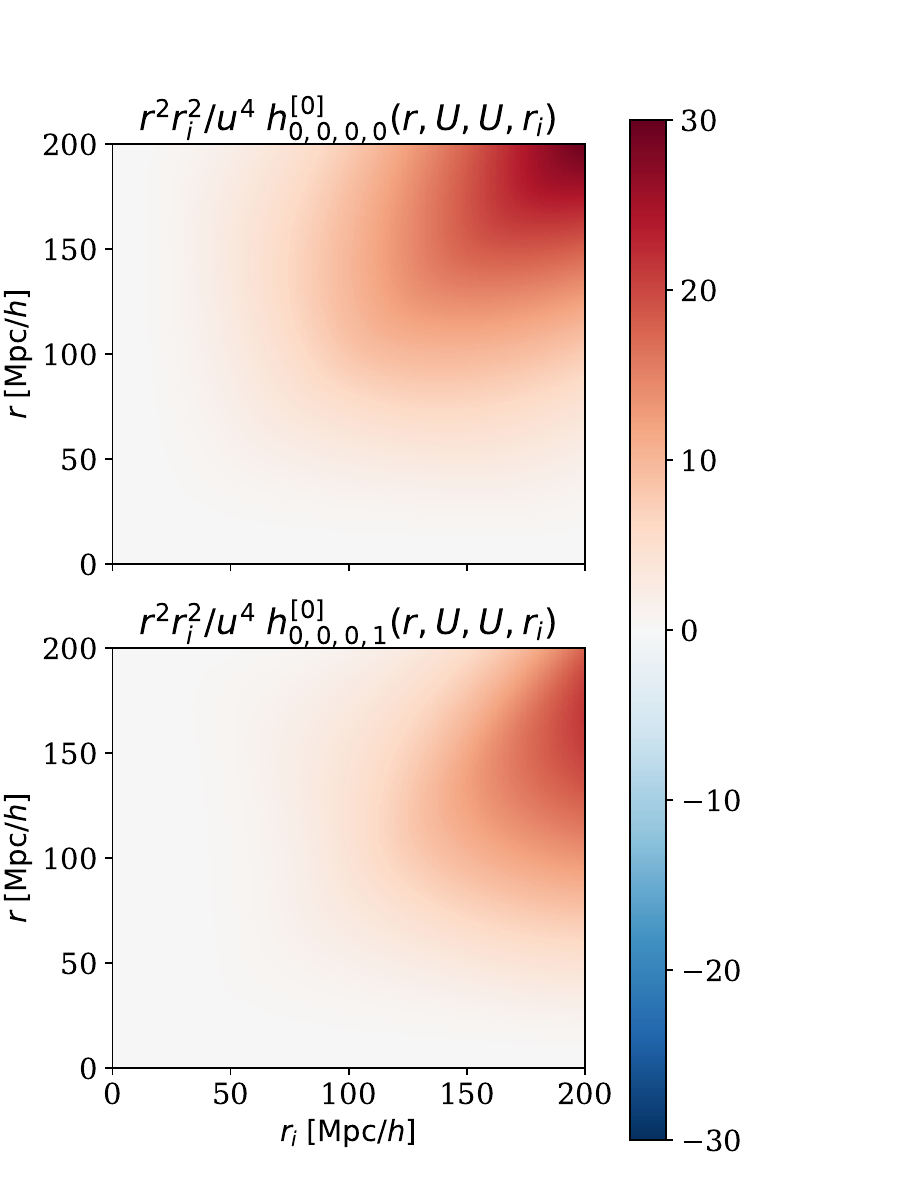}
\caption{Here, we show Eq. (\ref{eq:hint}) for fixed $r' = U \equiv 100 \;\left[{\rm Mpc}/h\right]$, $n' = 0$ $\ell'=0$,  $\ell = 0$ and $L = \left\{0,1\right\}$. The plot has been taken from section 4 of \cite{Ortola_4PCF}. The \textit{upper panel} shows the integral for $L=0$, while the \textit{lower panel} shows the integral for $L=1$. Since the 4PCF can be approximated as the square of the 2PCF on large scales, $(\xi_{0})^{2}(r) \sim (1/r^2)^2$, we have weighted the integral by $r^2r_i^2/u^{4}$, with $u \equiv 10 \;\left[{\rm Mpc}/h\right]$, to take out its fall-off. Both the \textit{upper} and \textit{lower} panel show that the behavior of the integral creates a rectangular boundary. Analytical results with the power spectrum using a power law and explaining the blobs that appear at about $r=r_i\approx U$ have been obtained in \cite{Ortola_4PCF}.}
\label{fig:hint}
\end{figure}

We proceed by writing the complex exponential in terms of the isotropic basis functions, which will allow us to separate the angular and radial parts completely; evaluating the $\vecq_3$ integral will follow the same procedure, and its result is explicitly shown in Eq. (\ref{eq:I_SS_q3}). Using the plane wave expansion (PWE) we find \cite{Ortola_4PCF}:
\begin{align}\label{eq:PWE}
&e^{-i \mathbf{q}_2 \cdot\left(\mathbf{r}_0-\mathbf{s}-\vecr'_1\right)} = (4\pi)^{3}\sum_{L_{q20},L_{q2s},L'_{q21}=0}^{\infty} C_{L_{q20},L_{q2s},L'_{q21}} \Upsilon_{L_{q20},L_{q2s},L'_{q21}} \; j_{L_{q20}}(q_2 r_2)  j_{L_{q2s}}(k_2 s)  \nonumber \\
& \qquad \qquad   \times j_{L'_{q21}}(k_2 r'_1) \mathcal{P}_{L_{q20},L_{q2s},L'_{q21}}(\hatq_2,\hatq_2,\hatq_2) \mathcal{P}_{L_{q20},L_{q2s},L'_{q21}}(\hatr_0,-\hats,-\hatr'_1). 
\end{align}
We define the PWE constants as:
\begin{align} \label{eq:Constant_from_exp}
C_{\ell'_1,\ell'_2,\ell'_3} \equiv i^{\ell'_1+\ell'_2+\ell'_3} \; \sqrt{(2\ell'_1+1)(2\ell'_2+1)(2\ell'_3+1)}, 
 \end{align}
and
\begin{align}\label{eq:Upsilon_def}
 \Upsilon_{L_1,L_2,L_3} =  \quad \frac{(-1)^{L_1+L_2+L_3}}{\sqrt{(2L_1+1)(2L_2+1)(2L_3+1)}}, 
\end{align}.
Inserting the above result into Eq. (\ref{eq:Def_of_I_2}) and evaluating the $\hatq_2$ integral we find:
\begin{align}\label{eq:I_SS_q2}
&I^{(q_2)}(\mathbf{r}_0,-\mathbf{s},-\vecr'_1,r) = (2\pi^2)(4\pi)^{3}  \sum_{L_{q20},L_{q2s},L'_{q21},\ell_2} C_{L_{q20},L_{q2s},L'_{q21}} \Upsilon_{L_{q20},L_{q2s},L'_{q21}} \mathcal{L}_{L_{q20},L_{q2s},L'_{q21}}^{j_{12},j_{23},j, \ell_2, \{m\}}  \nonumber \\
& \qquad \qquad \qquad \qquad \qquad  \times h^{[n_2'+n]}_{L_{q20},L_{q2s},L'_{q21},\ell_2}(r_0,s,r'_1,r) \mathcal{P}_{L_{q20},L_{q2s},L'_{q21}}(\hatr_0,-\hats,-\hatr'_1),
\end{align}
where the constant $\mathcal{L}_{L_{q20},L_{q2s},L'_{q21}}^{j_{12},j_{23},j,\ell_2,\{m\}}$ comes from evaluating the angular integral as shown in Eq. (\ref{eq:Fancy_L_def}). The radial integral $h$ is defined as:
\begin{align}\label{eq:hint}
h_{L, L',L^{''},L_i}^{[n]}(r, r', r^{''},r_i) \equiv \int \frac{dk}{2\pi^2} \;k^{n+2}j_{L}(k r) j_{L'}(k r') j_{L^{''}}(k r^{''}) j_{L_i}(k r_i) \;\Pk(k),
\end{align}
which we numerically compute in Fig. \ref{fig:hint} for the two lowest order angular momenta ($L=\{0,1\}$). We refer the reader to Appendix J of \cite{Ortola_4PCF}, where a more detailed analysis of the behavior of this radial integral is provided.
\begin{figure}[h!]
\centering
\includegraphics[scale=0.6]{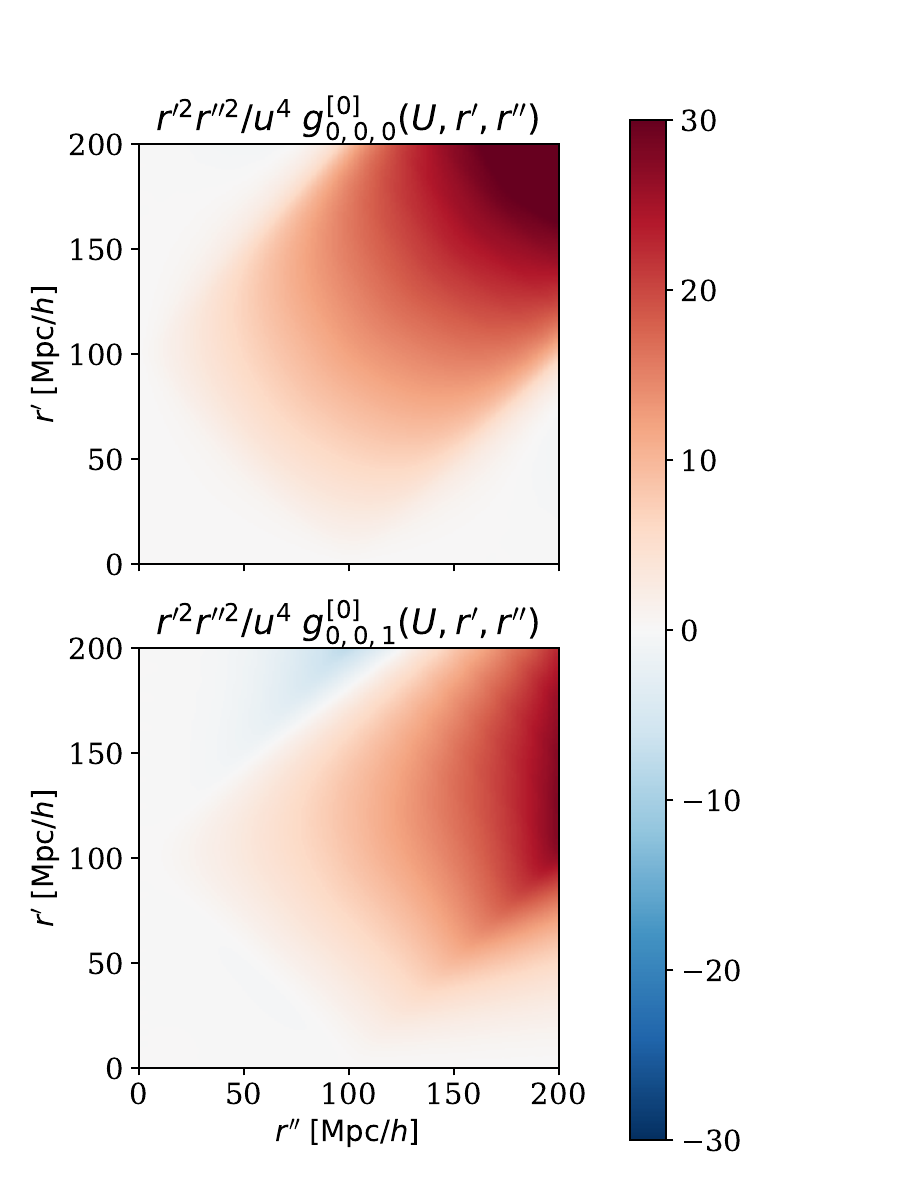}
\caption{Here, we show Eq. (\ref{eq:gint}) for fixed $r' = U \equiv 100 \;\left[{\rm Mpc}/h\right]$, $n' = 0$ $\ell'=0$,  $\ell = 0$ and $L = \left\{0,1\right\}$. The plot has been taken from section 3 of \cite{Ortola_4PCF}. The \textit{upper panel} shows the integral for $L=0$, while the \textit{lower panel} shows the integral for $L=1$. Since the 4PCF can be approximated as the square of the 2PCF on large scales, $(\xi_{0})^{2}(r) \sim (1/r^2)^2$, we have weighted the integral by $r^2r_i^2/u^{4}$, with $u \equiv 10 \;\left[{\rm Mpc}/h\right]$, to take out its fall-off. Both the \textit{upper} and \textit{lower} panel show that the behavior of the integral creates a rectangular boundary. Analytical results with the power spectrum using a power law and explaining the rectangular boundaries ($r' = r^"+U$, $r' = -r^"+U$, and $r' = r^"-U$) have been obtained in \cite{Ortola_4PCF}; further results are also evaluated in \cite{Chellino_3SBF}.}
\label{fig:gint2}
\end{figure}

Following the same steps yields the $\vecq_3$ integral as:
\begin{align}\label{eq:I_SS_q3}
&I^{(q_3)}(\mathbf{r}_0,-\mathbf{s},-\vecr'_2,r) = (2\pi^2)(4\pi)^{3} \sum_{L_{q30},L_{q3s},L'_{q3},\ell_2=0}^{\infty} C_{L_{q30},L_{q3s},L'_{q32}} \Upsilon_{L_{q30},L_{q3s},L'_{q32}} \mathcal{L}_{L_{q30},L_{q3s},L'_{q32}}^{j_{12},j_{23},j, \ell_2,\{m\}}  \nonumber \\
& \qquad \qquad \qquad \qquad \qquad  \times h^{[n_3'+n]}_{L_{q30},L_{q3s},L'_{q32},\ell_2}(r_0,s,r'_2,r) \mathcal{P}_{L_{q30},L_{q3s},L'_{q32}}(\hatr_0,-\hats,-\hatr'_2).
\end{align}
  
Next, we must evaluate the $\vecq_1$ integral. Doing so will follow the same procedure as before but without the coupled wave vector in the denominator; we must compute:
\begin{align}
I^{(q_1)}(\vecr_0,-\vecs,-\vecr'_0) = \int_{\vecq_1}q_1^{n'_1}\,Y_{j_{12},m_{j_{12}}}(\hatq_1)Y_{j_{13},m_{j_{13}}}(\hatq_1) P\left(q_1\right) \; e^{-i \vecq_1 \cdot\left(\mathbf{r}_0-\mathbf{s}-\vecr'_0\right)},
\end{align}
where we can expand the complex exponential in terms of the isotropic basis functions as in Eq. (\ref{eq:PWE}) and evaluate the angular integral with Eq. (\ref{eq:Fancy_J_def}) to obtain:
\begin{align}\label{eq:I_SS_1}
&I^{(q_1)}(\vecr_0,-\vecs,-\vecr'_0) =(2\pi^2)(4\pi)^{3} \sum_{L_{q10},L_{q1s},L'_{q10}=0}^{\infty} C_{L_{q10},L_{q1s},L'_{q10}} \Upsilon_{L_{q10},L_{q1s},L'_{q10}} \mathcal{J}_{L_{q10},L_{q1s},L'_{q10}}^{j_{12},j_{13},m_{j_{12}},m_{j_{13}}}  \nonumber \\
& \qquad \qquad \qquad \qquad \qquad  \times g^{[n_1']}_{L_{q10},L_{q1s},L'_{q10}}(r_0,s,r'_0) \mathcal{P}_{L_{q10},L_{q1s},L'_{q10}}(\hatr_0,-\hats,-\hatr'_0).
\end{align}
The radial integral $g$ is defined as:
\begin{align}\label{eq:gint}
g_{ L,L',L''}^{[n]}(r, r',r'') \equiv \int \frac{dk}{2\pi^2} \;k^{n+2} j_{L}(k r) j_{L'}(k r') j_{L''}(k r'') \;\Pk(k),
\end{align}
which we numerically compute in Fig. \ref{fig:gint2} for the two lowest order angular momenta ($L=\{0,1\}$). We refer the reader to Appendix J of \cite{Ortola_4PCF}, where a more detailed analysis of the behavior of this radial integral is provided.

We now compute the integral of $\veck_3$. We observe in Eq. (\ref{eq:middlestep_SS_config}), that $\veck_3$ is not affected by the third-order kernel:
\begin{align}
I^{(k_3)}(\vecr_3,-\vecs,-\vecr'_3) = \int_{\veck_3} P\left(k_3\right) \; e^{-i \veck_3 \cdot\left(\mathbf{r}_3-\mathbf{s}-\vecr'_3\right)}.
\end{align}
The only angular dependence for performing the $\hatk_3$ integral comes from the exponential, which we evaluate in Eq. (\ref{eq:modified_Gaunt}), and we can immediately find:
\begin{align}\label{eq:I_SS_k3}
&I^{(k_3)}(\vecr_3,-\vecs,-\vecr'_3) =(4\pi)^2 \sum_{L_{33},L'_{3s},L'_{33} = 0}^{\infty} C_{L_{33},L'_{3s},L'_{33}} \Upsilon_{L_{33},L'_{3s},L'_{33}} \mathcal{G}_{L_{33},L'_{3s},L'_{33}}  \nonumber \\
& \qquad \qquad \qquad \qquad \qquad  \times g^{[0]}_{L_{33},L'_{3s},L'_{33}}(r_3,s,r'_3) \mathcal{P}_{L_{33},L'_{3s},L'_{33}}(\hatr_3,-\hats,-\hatr'_3).  
\end{align} 

\begin{figure}[h!]
\centering
\includegraphics[scale=0.4]{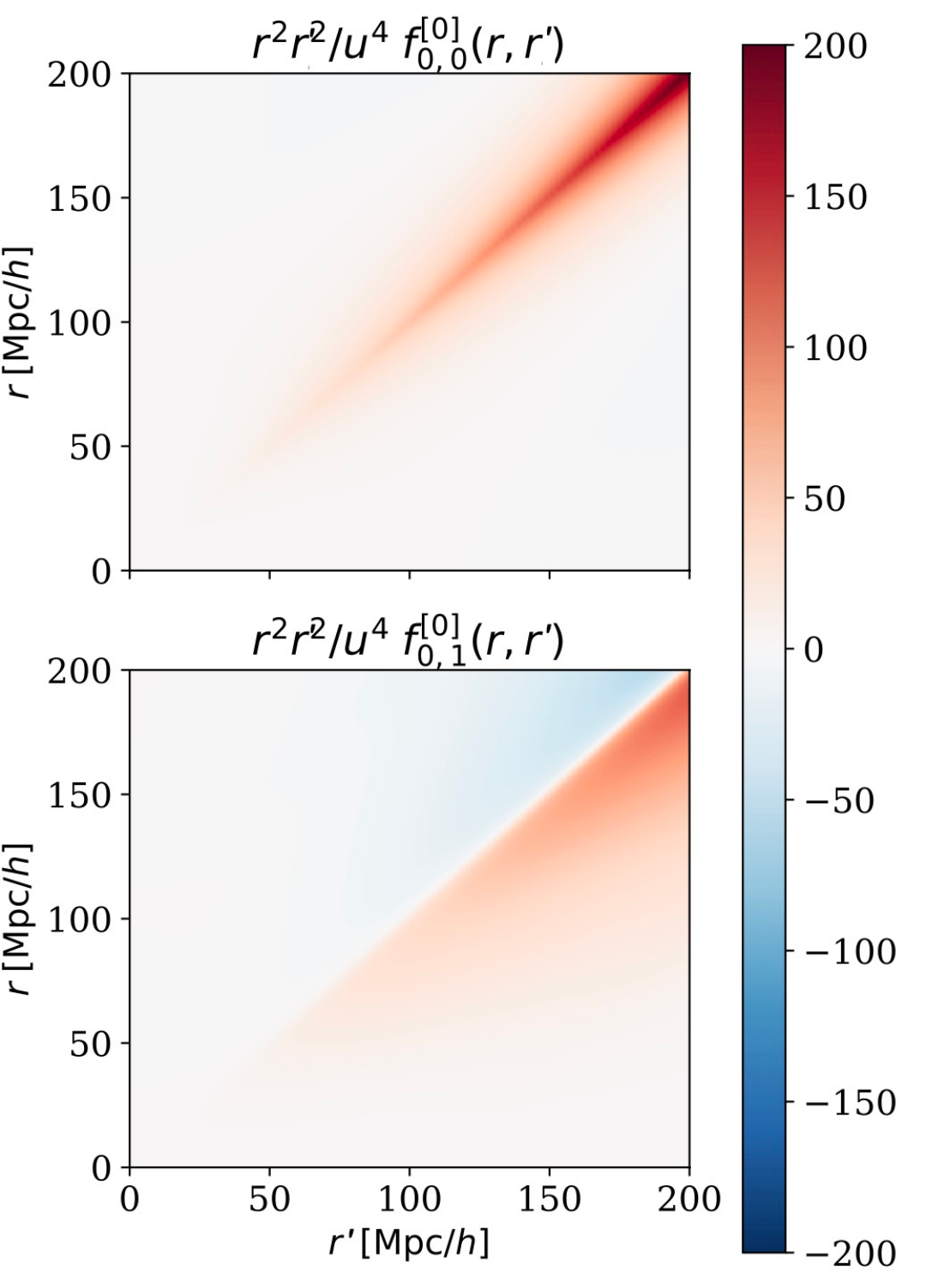}
\caption{Here, we show Eq. (\ref{eq:f_radial_def}) choosing $n = 0$, $\ell=0$, and $\ell' = \left\{0,1\right\}$. The plot has been taken from section 3 of \cite{Ortola_4PCF}. The \textit{upper panel} shows the integral for $\ell'=0$ and is largest along the diagonal. The \textit{lower panel} shows the integral for $\ell'=1$ and is largest in the off-diagonal elements. Since the 4PCF can be approximated as the square of the 2PCF on large scales, $(\xi_{0})^{2}(r) \sim (1/r^2)^2$, we have weighted the integral by $r^2r_i^2/u^{4}$, with $u \equiv 10 \;\left[{\rm Mpc}/h\right]$, to take out its fall-off.}
\label{fig:f_int}
\end{figure}

\begin{figure}[h!]
\centering
\includegraphics[scale=0.6]{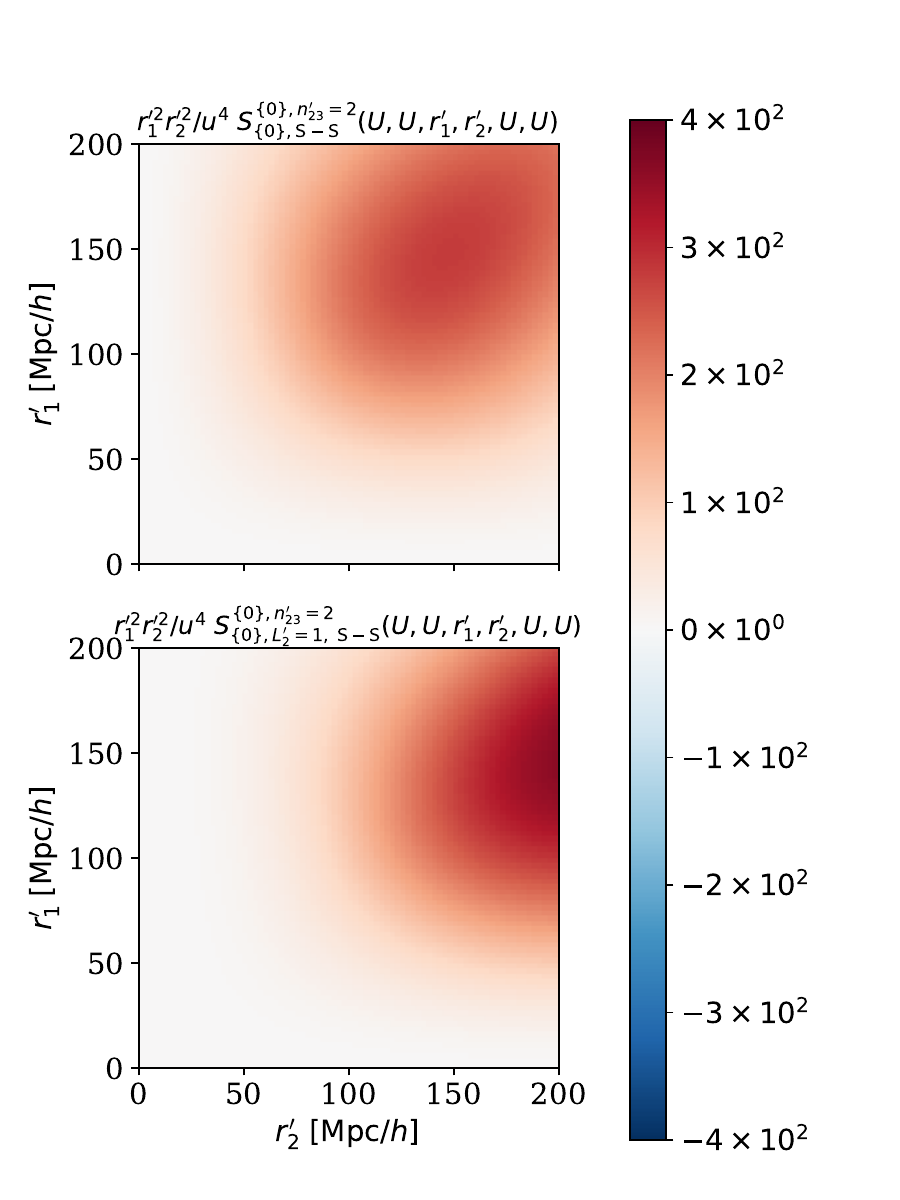}
\caption{Here, we show Eq. (\ref{eq:Sint_SS}) with fixed arguments at $ U \equiv 100 \;\left[{\rm Mpc}/h\right]$, except for $r'_1$ and $r'_2$. We also chose to evaluate $n' = n=0$, $n'_{23}=2$ and all angular momenta $\{L\} = 0$, except for $L'_2 = \left\{0,1\right\}$. The \textit{upper panel} shows the integral for $L'_2=0$, while the \textit{lower panel} shows the integral for $L'_2=1$. Since the 4PCF can be approximated as the square of the 2PCF on large scales, $(\xi_{0})^{2}(r) \sim (1/r^2)^2$, we have weighted the integral by $r_1^{'2}r_2^{'2}/u^{4}$, with $u \equiv 10 \;\left[{\rm Mpc}/h\right]$, to take out its fall-off. The $S$ integral is composed of the intermediate radial integrals, $g$ and $h$, which reveal a rectangular boundary as shown in Fig. \ref{fig:gint2}, and a blob appearing at approximately the chosen value of $U$ as shown in Fig. \ref{fig:hint}. The integration of the different rectangular boundaries and the blob creates the oval-shaped regions shown in both panels. Analytical results for similar radial integral with the power spectrum using a power law have been obtained in \cite{Ortola_4PCF}.}
\label{fig:S_SS_int}
\end{figure}

The next integral to evaluate is that over $\veck_1$. Analogously to the $\veck_3$ integral, there is no contribution from the third-order kernel and the angular dependence comes from the complex exponential. Since this latter's argument contains only two vectors, we evaluate it explicitly here:
\begin{align}\label{eq:PWE_2args}
&e^{-i \mathbf{k}_1 \cdot\left(\mathbf{r}_1-\vecr_2\right)} = (4\pi)^{2}\sum_{L_{11},L_{22}=0}^{\infty}\sum_{M_{11}=-L_{11}}^{L_{11}}\sum_{M_{22}=-L_{22}}^{L_{22}} i^{L_{22}-L_{11}} \; j_{L_{11}}(k_1 r_1)  j_{L_{22}}(k_1 r_2) \nonumber \\
& \qquad \qquad \qquad \qquad \quad \times Y_{L_{11}, M_{11}}(\hatk_1)Y_{L_{22}, M_{22}}^{*}(\hatk_1)Y_{L_{11}, M_{11}}^*(\hatr_1)Y_{L_{22}, M_{22}}(\hatr_2),  
\end{align}
where we use the spherical harmonics basis. The angular integral will only be composed of these two spherical harmonics, resulting in the product of Kronecker deltas, $\dK_{L_{11},L_{22}}\dK_{M_{11},M_{22}}$, allowing us to obtain in terms of the isotropic basis functions:
\begin{align}\label{eq:def_I_k1}
&I^{(k_1)}(\vecr_1,\vecr_2) =(4\pi) \sum_{L_{11}}(-1)^{L_{11}}s_{L_{11}}^{(\rm I)} f^{[0]}_{L_{11},L_{11}}(r_1,r_2) \mathcal{P}_{L_{11}}(\hatr_1,\hatr_2),
\end{align}
where $s_{L_{11}}^{(\rm I)}\equiv \sqrt{2L_{11}+1}$ and the radial integral has been defined as:
\begin{align}\label{eq:f_radial_def}
f^{[n]}_{L_{1},L_{2}}(r_1,r_2) = \int \frac{dk}{2\pi^2}\;k^{n+2}\;j_{L_1}(kr_1)j_{L_2}(kr_2)\;P(k),
\end{align}
which we numerically compute in Fig. \ref{fig:f_int} for the first two lowest order angular momenta ($L=\{0,1\}$).
The final angular dependence, on $\hats$, we evaluate in $\S$\ref{sec:s_hat_SS_Evaluation}, leading to the complete expression for the secondary-secondary contribution:
\begin{empheq}[box=\widefbox]{align}\label{eq:SS_res}
&{\rm Cov}_{4, \rm 1L}^{[3],{\rm S-S}} (\mathbf{R},\mathbf{R}') = (4\pi)^{12} \pi^3 \sum_{\rm All} \nonumber \\
& \qquad \qquad \times (-1)^{\Lambda+\Lambda^{(4)}+\ell_2}\; \mathcal{Q}^{\Lambda,\Lambda_s,\Lambda'}\left(s_{\Lambda^{(4)}}^{(\rm I)}\right)^{-1} \nonumber \\
& \qquad \qquad \times c_{j,n}^{(\mu)} d_{j_{12}}^{(n_{12})}d_{j_{13}}^{(n_{13})}d_{j_{23}}^{(n_{23})}\;D_j \; s_{L_{11}}^{(\rm I)} \Omega_{n'_{23}}\mathcal{D}_{n'_1,n'_2,n'_3,n'_{23}}^{n_{12},n_{13},n_{23}}
\nonumber \\
& \qquad \qquad \times C_{L_{q10},L_{q1s},L'_{q10}} \Upsilon_{L_{q10},L_{q1s},L'_{q10}} \mathcal{J}_{L_{q10},L_{q1s},L'_{q10}}^{j_{12},j_{13},m_{j_{12}},m_{j_{13}}} \nonumber \\
& \qquad \qquad \times C_{L_{q20},L_{q2s},L'_{q21}} \Upsilon_{L_{q20},L_{q2s},L'_{q21}} \mathcal{L}_{L_{q20},L_{q2s},L'_{q21}}^{j_{12},j_{23},j, \ell_2,\{m\}}\nonumber \\
& \qquad \qquad \times  C_{L_{q30},L_{q3s},L'_{q32}} \Upsilon_{L_{q30},L_{q3s},L'_{q32}} \mathcal{L}_{L_{q30},L_{q3s},L'_{q32}}^{j_{12},j_{23},j, \ell_2,\{m\}} \nonumber \\,
& \qquad \qquad \times C_{L_{33},L'_{33},L'_{3s}} \Upsilon_{L_{33},L'_{33},L'_{3s}} \mathcal{G}_{L_{33},L'_{33},L'_{3s}} \nonumber \\
& \qquad \qquad \times f_{L_{11},L_{11}}^{[0]}(r_1,r_2) S_{\{L\}, {\rm S-S}}^{(\{L_{is}\})}(r_0,r'_0,r'_1,r'_2,r_3,r'_3) \nonumber \\
& \qquad \qquad \times \PP_{\Lambda}(\hatR)\PP_{\Lambda'}(\hatR')+\;71\;{\rm perms.}.
\end{empheq}
In the sum above, ``All'' denotes the sum over all angular momentum and magnetic quantum numbers appearing in the expression. The summation range for each angular momentum variable is defined in the equation where that variable is first introduced. For convenience, Table~\ref{table:1} lists all angular momentum coupling coefficients together with the equations in which they are defined. We have also defined the radial integral $S$ as:
\begin{align}\label{eq:Sint_SS}
&S_{\{L\}, {\rm S-S}}^{(\{L_{is,r}\})}(r_0,r'_0,r'_1,r'_2,r_3,r'_3) \nonumber \\
& \qquad  \equiv \int dr \;r^{n'_{23}-1}\int ds\;s^2\; g_{L_{q10},L'_{1qs},L'_{q10}}^{[0]} (r_0,s,r'_0) h_{L_{q20},L_{q2s},L'_{q21},\ell_2}^{[n+n'_2]} (r_0,s,r'_{1},r)\nonumber \\
& \qquad \quad \times h_{L_{q30},L_{q3s},L'_{q32},\ell_2}^{[n+n'_3]} (r_0,s,r'_{2},r)
g_{L_{33},L'_{3s},L'_{33}}^{[0]} (r_3,s,r'_{3}).
\end{align}
The subscript $\{L\}$ denotes the set of all angular momentum indices that affect one of the position-space variables (\textit{i.e.}, the $r_0$, $r'_0$,$\cdots$ variables). The superscript $(\{L_{is,r}\})$ denotes the set of all angular momentum indices that are coupled to $s$ and $r$ in the spherical Bessel functions. Numerical evaluation of this radial integral for the lowest order angular momenta ($L=\{0,1\}$) is shown in Fig. \ref{fig:S_SS_int}. The behavior of this radial integral is smooth and predominantly positive over the plotted domain. For fixed \(U=100\,\mathrm{Mpc}/h\), it is suppressed for \(r_1',r_2' \ll U\) and grows when the primed separations become comparable to or exceed \(U\), indicating sensitivity to relative scale matching rather than an intrinsic large-scale enhancement. Its net effect on the full NLO 4PCF covariance must nevertheless be assessed only after including the remaining structures, angular couplings, and permutations.

\subsection{Primary-Secondary Configuration}\label{sec:P-S_conf}
Using Wick's theorem for the primary-secondary configuration shown in Fig. \ref{fig:dc_configurations} on Eq. (\ref{eq:sub_cov_before_Wicks}) results in:
\begin{align}
&{\rm Cov}_{4, \rm 1L}^{[3],{\rm P-S}}(\mathbf{R},\mathbf{R}') = \prod_{v=0}^{3} \int_{\vecs}\int_{\veck_v}\int_{\veck'_v}\left[ \;e^{-i\veck_v\cdot(\vecx+\vecr_v)} e^{-i\veck'_v \cdot(\vecx+\vecr'_v+\vecs)} \right]\nonumber \\
& \qquad \qquad \qquad \times\int d^3\vecq_1 \int d^3\vecq_2 \int d^3\vecq_3 \ \;\W(\vecq_1,\vecq_2,\vecq_3) \; \dD(\veck_0-\vecq_1-\vecq_2-\vecq_3)\nonumber \\
& \qquad \qquad \qquad \times  \left<\widetilde{\delta}_{\rm lin.}(\vecq_1)\widetilde{\delta}_{\rm lin.}(\veck_1)\right>\left<\widetilde{\delta}_{\rm lin.}(\vecq_2)\widetilde{\delta}_{\rm lin.}(\veck'_0)\right> \left<\widetilde{\delta}_{\rm lin.}(\vecq_3)\widetilde{\delta}_{\rm lin.}(\veck'_1)\right>\nonumber \\
& \qquad \qquad \qquad \times\left<\widetilde{\delta}_{\rm lin.}(\veck_2)\widetilde{\delta}_{\rm lin.}(\veck'_2)\right> \left<\widetilde{\delta}_{\rm lin.}(\veck_3)\widetilde{\delta}_{\rm lin.}(\veck'_3)\right> + 71\; {\rm perms.},
\end{align}
where $71$ perms. indicates the other 71 combinations\footnote{Finding the number of permutations is done in four steps. i) Start with one of the third-order densities, then pair this third-order density with one of the linear densities from the other tetrahedron; one has 4 possibilities. ii) Repeat with one the two remaining third-order densities; one has 3 possibilities. iii) Pair the remaining third-order density with one of the linear densities in the same tetrahedron; one has 3 possibilities. iv) Pair the remaining two linear densities in each tetrahedron; one has 2 possibilities Total number of permutations = 4 $\times$ 3$\times$ 3 $\times$ 2 = 72 perms.} of Wick's theorem that we can obtain for this Primary-Secondary structure. For brevity, we leave this out moving forward, but it is implicitly included into our calculations. Applying Wick's theorem and evaluating the integrals over $\veck_0,\;\veck_1,\;\veck'_0,\;\veck'_1,\;\veck'_2,\;{\rm and}\;\veck'_3$, we obtain:
\begin{align}
&{\rm Cov}_{4, \rm 1L}^{[3],{\rm P-S}}(\mathbf{R},\mathbf{R}') = \int_{\vecs,\; \veck_2,\;\veck_3,\;\vecq_1,\;\vecq_2,\;\vecq_3} e^{-i\vecq_1\cdot(\vecr_0-\vecr_1)} e^{-i\vecq_2\cdot(\vecr_0-\vecr'_0-\vecs)} \nonumber\\
& \qquad \qquad \qquad \qquad  \times e^{-i\vecq_3\cdot(\vecr_0-\vecr'_1-\vecs)} e^{-i\veck_2\cdot(\vecr_2-\vecr'_2-\vecs)}e^{-i\veck_3\cdot(\vecr_3-\vecr'_3-\vecs)} \nonumber\\
& \qquad \qquad \qquad \qquad \times P(q_1) P(q_2) P(q_3) P(k_2) P(k_3) \W(\vecq_1,\vecq_2,\vecq_3).
\end{align}
We have already solved the integrals over $\vecq_2\;{\rm and}\;\vecq_3$ in Eqs. (\ref{eq:I_SS_q2}), (\ref{eq:I_SS_q3}), and the $\;\veck_2,\; {\rm and}\;\veck_3$ in Eq. (\ref{eq:I_SS_k3}). We write them out explicitly again here since they will have different angular momenta dependence:
\begin{align}\label{eq:I_PS_q2}
&I^{(q_2)}(\mathbf{r}_0,-\mathbf{s},-\vecr'_0,r) = (2\pi^2)(4\pi)^{3} \sum_{L_{q20},L_{q2s},L'_{q20},\ell_2} C_{L_{q20},L_{q2s},L'_{q20}} \Upsilon_{L_{q20},L_{q2s},L'_{q20}} \mathcal{L}_{L_{q20},L_{q2s},L'_{q20}}^{j_{12},j_{23},j,\ell_2,\{m\}}  \nonumber \\
& \qquad \qquad \qquad \qquad \qquad  \times h^{[n_2'+n]}_{L_{q20},L_{q2s},L'_{q20},\ell_2}(r_0,s,r'_0,r) \mathcal{P}_{L_{q20},L_{q2s},L'_{q20}}(\hatr_0,-\hats,-\hatr'_0),
\end{align}
\begin{align}\label{eq:I_PS_q3}
&I^{(q_3)}(\mathbf{r}_0,-\mathbf{s},-\vecr'_3,r) = (2\pi^2)(4\pi)^{3} \sum_{L_{q30},L_{q3s},L'_{q31},\ell_2} C_{L_{q30},L_{q3s},L'_{q31}} \Upsilon_{L_{q30},L_{q3s},L'_{q31}} \mathcal{L}_{L_{q30},L_{q3s},L'_{q31}}^{j_{13},j_{23},j,\ell_2,\{m\}}  \nonumber \\
& \qquad \qquad \qquad \qquad \qquad  \times h^{[n_3'+n]}_{L_{q30},L_{q3s},L'_{q31},\ell_2}(r_0,s,r'_3,r) \mathcal{P}_{L_{q30},L_{q3s},L'_{q31}}(\hatr_0,-\hats,-\hatr'_1),
\end{align}
\begin{align}\label{eq:I_PS_k2}
&I^{(k_2)}(\vecr_2,-\vecs,-\vecr'_2) =(4\pi)^2 \sum_{L_{22},L'_{2s},L'_{22}} C_{L_{22},L'_{2s},L'_{22}} \Upsilon_{L_{22},L'_{2s},L'_{22}} \mathcal{G}_{L_{22},L'_{2s},L'_{22}}  \nonumber \\
& \qquad \qquad \qquad \qquad \qquad  \times g^{[0]}_{L_{22},L'_{2s},L'_{22}}(r_2,s,r'_2) \mathcal{P}_{L_{22},L'_{2s},L'_{22}}(\hatr_2,-\hats,-\hatr'_2),    
\end{align} 
\begin{align}\label{eq:I_PS_k3}
&I^{(k_3)}(\vecr_3,-\vecs,-\vecr'_1) =(4\pi)^2 \sum_{L_{33},L'_{3s},L'_{31}} C_{L_{33},L'_{3s},L'_{31}} \Upsilon_{L_{33},L'_{3s},L'_{31}} \mathcal{G}_{L_{33},L'_{3s},L'_{31}}  \nonumber \\
& \qquad \qquad \qquad \qquad \qquad  \times g^{[0]}_{L_{33},L'_{3s},L'_{31}}(r_3,s,r'_1) \mathcal{P}_{L_{33},L'_{3s},L'_{31}}(\hatr_3,-\hats,-\hatr'_1),    
\end{align} 
where the summation range for each of these angular momentum variables goes from zero to infinity as already shown in $\S$\ref{sec:S-S_conf}. We remind the reader the sums from the $W^{(3)}$ kernel are omitted above since they are used across multiple equations, but we introduce them back in Eq. (\ref{eq:PS_res}).

\begin{figure}[h!]
\centering
\includegraphics[scale=0.6]{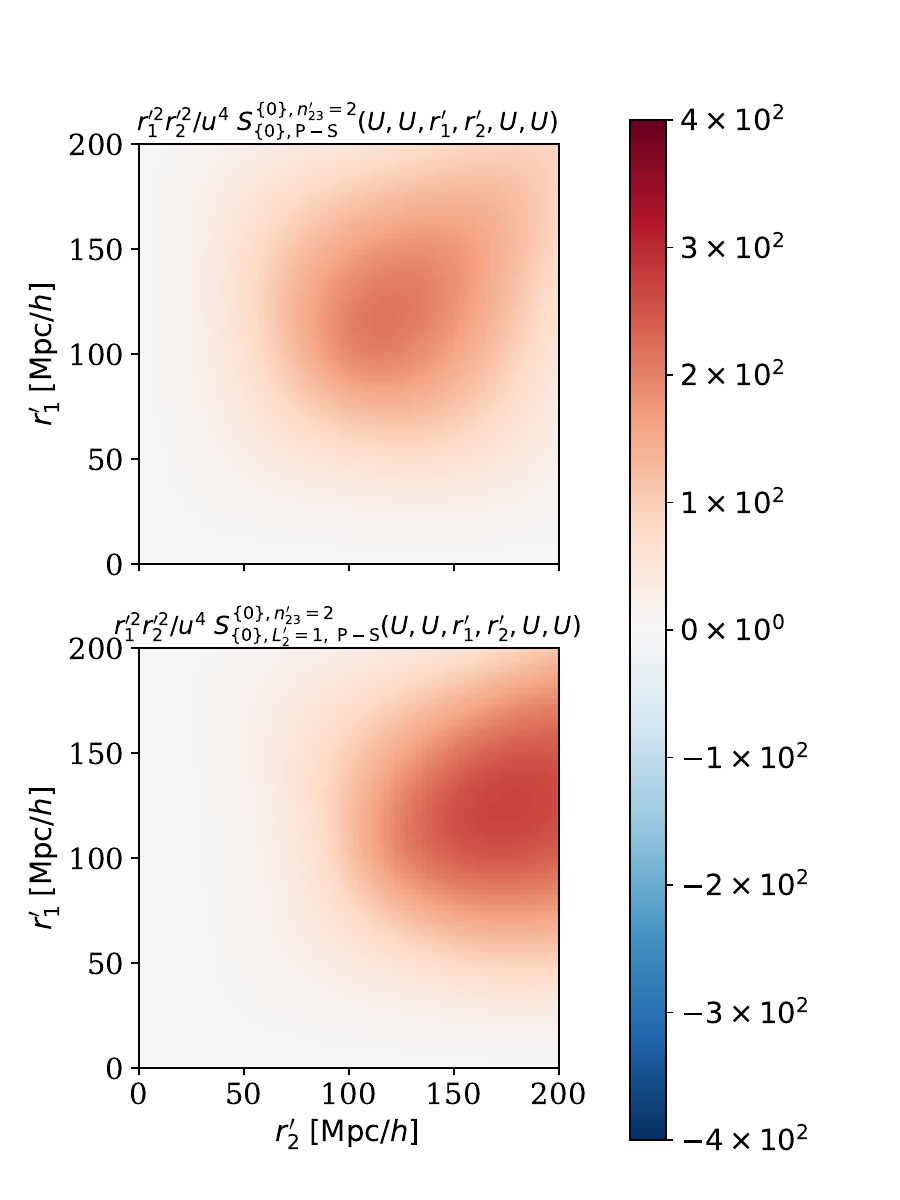}
\caption{Here, we show Eq. (\ref{eq:Sint_PS}) with fixed arguments at $ U \equiv 100 \;\left[{\rm Mpc}/h\right]$, except for $r'_1$ and $r'_2$. We also chose to evaluate $n' = n=0$, $n'_{23}=2$ and all angular momenta $\{L\} = 0$, except for $L'_2 = \left\{0,1\right\}$. The \textit{upper panel} shows the integral for $L'_2=0$, while the \textit{lower panel} shows the integral for $L'_2=1$. Since the 4PCF can be approximated as the square of the 2PCF on large scales, $(\xi_{0})^{2}(r) \sim (1/r^2)^2$, we have weighted the integral by $r_1^{'2}r_2^{'2}/u^{4}$, with $u \equiv 10 \;\left[{\rm Mpc}/h\right]$, to take out its fall-off. The $S$ integral is composed of the intermediate radial integrals, $g$ and $h$, which reveal a rectangular boundary as shown in Fig. \ref{fig:gint2}, and a blob appearing at approximately the chosen value of $U$ as shown in Fig. \ref{fig:hint}. The integration of the different rectangular boundaries and the blob creates the oval-shaped regions shown in both panels. Analytical results for similar radial integral with the power spectrum using a power law have been obtained in \cite{Ortola_4PCF}.}
\label{fig:S_PS_int}
\end{figure}

Next, we must evaluate the $\vecq_1$ integral since it is the only one to have a different result:
\begin{align}
I^{(q_1)}(\vecr_0,\vecr_1) = \int_{\vecq_1} e^{-i\vecq_1\cdot(\vecr_0-\vecr_1)\;} q^{n_1'} \;Y_{j_{12},m_{j_{12}}}(\hatq_1)Y_{j_{13},m_{j_{13}}}(\hatq_1)\; P(q_1).
\end{align}
Expanding the complex exponential using the plane wave expansion, we can solve the angular integral with Eq. (\ref{eq:Def_H_bar}) and obtain:
\begin{align}\label{eq:I_PS_q1}
&I^{(q_1)}(\vecr_0,\vecr_1) = (2\pi^2)(4\pi)^{2}\;\sum_{L_{q10},L_{q11}}w_{L_{q10},L_{q11}}\overline{\mathcal{H}}_{L_{q10},L_{q11},M_{q10},-M_{q11}}^{j_{12},j_{13},m_{j_{12}},m_{j_{13}}}\nonumber\\
& \qquad \qquad \qquad \; \times f^{[n'_1]}_{L_{q10},L_{q11}}(r_0,r_1)Y_{L_{q10},M_{q10}}^*(\hatr_0)Y_{L_{q11},M_{q11}}(\hatr_1),
\end{align}
where we have defined $w$ as:
\begin{align}
    w_{L,L'}\equiv i^{L-L'}.
\end{align}
With the above result, the only remaining difficulty is the analysis of the $\hats$ integral, which we evaluate in $\S$\ref{sec:s_hat_PS_Evaluation}. Therefore, the Primary-Secondary configuration yields:
\begin{empheq}[box=\widefbox]{align}\label{eq:PS_res}
&{\rm Cov}_{4, \rm 1L}^{[3],{\rm P-S}} (\mathbf{R},\mathbf{R}') = (2\pi^2)^{3} (4\pi)^{12} \sum_{\rm All}  \nonumber \\
& \qquad \qquad \times \overline{\mathcal{Q}}_{(\rm P-S)}^{\Lambda,\Lambda_s,\Lambda'}  d_{j_{12}}^{(n_{12})}d_{j_{13}}^{(n_{13})}d_{j_{23}}^{(n_{23})} D_j\;\mathcal{D}_{n'_1,n'_2,n'_3,n'_{23}}^{n_{12},n_{13},n_{23}} \nonumber \\
& \qquad \qquad \times \mathcal{D}_{n'_1,n'_2,n'_3,n'_{23}}^{n_{12},n_{13},n_{23}}\;w_{L_{q10},L_{q11}}\;\overline{\mathcal{H}}_{L_{q10},L_{q11},M_{q10},-M_{q11}}^{j_{12},j_{13},m_{j_{12}},m_{j_{13}}}  \nonumber \\
& \qquad \qquad \times C_{L_{q20},L_{q2s},L'_{q20}} \Upsilon_{L_{q20},L_{q2s},L'_{q20}} \mathcal{L}_{L_{q20},L_{q2s},L'_{q20}}^{j_{12},j_{23},j,\ell_2,\{m\}} \nonumber \\
& \qquad \qquad \times C_{L_{q30},L_{q3s},L'_{q31}} \Upsilon_{L_{q30},L_{q3s},L'_{q31}} \mathcal{L}_{L_{q30},L_{q3s},L'_{q31}}^{j_{13},j_{23},j,\ell_2,\{m\}}\nonumber \\
& \qquad \qquad \times C_{L_{22},L'_{2s},L'_{22}} \Upsilon_{L_{22},L'_{2s},L'_{22}} \mathcal{G}_{L_{22},L'_{2s},L'_{22}} \nonumber \\
& \qquad \qquad \times C_{L_{33},L'_{3s},L'_{31}} \Upsilon_{L_{33},L'_{3s},L'_{31}} \mathcal{G}_{L_{33},L'_{3s},L'_{31}}\nonumber \\
& \qquad \qquad \times S_{\{L\}, {\rm P-S}}^{(\{L_{is}\})}(r_0,r'_0,r_1,r'_1,r_2,r'_2,r_3,r'_3) \nonumber \\
& \qquad \qquad \times \PP_{\Lambda}(\hatR)\PP_{\Lambda'}(\hatR')+\;71\;{\rm perms.}.
\end{empheq}
In the sum above, ``All'' denotes the sum over all angular momentum and magnetic quantum numbers appearing in the expression. The summation range for each angular momentum variable is defined in the equation where that variable is first introduced. For convenience, Table~\ref{table:1} lists all angular momentum coupling coefficients together with the equations in which they are defined. We have also defined $S$ as:
\begin{align}\label{eq:Sint_PS}
&S_{\{L\}, {\rm P-S}}^{(\{L_{is}\})}(r_0,r'_0,r'_1,r_2,r'_2,r_3,r'_3) \nonumber \\
& \qquad  \equiv \int dr \;r^{n'_{23}-1}\int ds\;s^2\; h^{[n_2'+n]}_{L_{q20},L_{q2s},L'_{q20},\ell_2}(r_0,s,r'_0,r) h^{[n_3'+n]}_{L_{q30},L_{q3s},L'_{q31},\ell_2}(r_0,s,r'_3,r)\nonumber \\
& \qquad \qquad \qquad \qquad \quad \times  g^{[0]}_{L_{22},L'_{2s},L'_{22}}(r_2,s,r'_2) g^{[0]}_{L_{33},L'_{3s},L'_{31}}(r_3,s,r'_1).
\end{align} 
The subscript $\{L\}$ denotes the set of all angular momentum indices that affect one of the position-space variables (\textit{i.e.}, the $r_0$, $r'_0$,$\cdots$ variables). The superscript $(\{L_{is,r}\})$ denotes the set of all angular momentum indices that are coupled to $s$ and $r$ in the spherical Bessel functions. Numerical evaluation of this radial integral for the lowest order angular momenta ($L=\{0,1\}$) is shown in Fig. \ref{fig:S_PS_int}. This radial integral exhibits behavior very similar to that shown in Fig.~\ref{fig:S_SS_int}, and we therefore adopt the same interpretation.

\subsection{Primary-Primary Configuration}

\qquad Using Wick's theorem for the primary-primary configuration shown in Fig. \ref{fig:dc_configurations} on Eq. (\ref{eq:sub_cov_before_Wicks}) results in:
\begin{align}
&{\rm Cov}_{4, \rm 1L}^{[3],{\rm P-P}}(\mathbf{R},\mathbf{R}') = \prod_{v=0}^{3} \int_{\vecs}\int_{\veck_v}\int_{\veck'_j=v}\left[\;e^{-i\veck_v\cdot(\vecx+\vecr_v)} e^{-i\veck'_v \cdot(\vecx+\vecr'_v+\vecs)}\right]\nonumber \\
& \qquad \qquad \qquad \times \int d^3\vecq_1 \int d^3\vecq_2 \int d^3\vecq_3 \; \W(\vecq_1,\vecq_2,\vecq_3)  \; \dD(\veck_0-\vecq_1-\vecq_2-\vecq_3)\nonumber \\
& \qquad \qquad \qquad \times  \left<\widetilde{\delta}_{\rm lin.}(\vecq_1)\widetilde{\delta}_{\rm lin.}(\vecq_2)\right>\left<\widetilde{\delta}_{\rm lin.}(\vecq_3)\widetilde{\delta}_{\rm lin.}(\veck'_0)\right> \left<\widetilde{\delta}_{\rm lin.}(\veck_1)\widetilde{\delta}_{\rm lin.}(\veck'_1)\right>\nonumber \\
& \qquad \qquad \qquad \times\left<\widetilde{\delta}_{\rm lin.}(\veck_2)\widetilde{\delta}_{\rm lin.}(\veck'_2)\right> \left<\widetilde{\delta}_{\rm lin.}(\veck_3)\widetilde{\delta}_{\rm lin.}(\veck'_3)\right> + 71{\rm \; perms.},
\end{align}
 where $71$ perms. indicates the other 71 combinations\footnote{Finding the number of permutations is done in three steps. i) Pair two of the third-order densities; one has 3 possibilities. ii) Pair the remaining third-order density with a linear density from the other tetrahedron; one has 4 possibilities. iii) Pair the remaining three linear densities in each tetrahedron; one has 6 possibilities. Total number of permutations = 3 $\times$ 4$\times$ 6  = 72 perms.} of Wick's theorem that we can obtain for this Primary-Primary structure. For brevity, we leave this out, but it is implicitly included into our calculations.
 The evaluation of the $W^{(3)}$ kernel, as shown in Eq. (\ref{eq:W_3_def}), also involves the $W^{(2)}$ kernel. The second-order kernel vanishes when $W^{(2)}(\vecq_i,-\vecq_i) = 0$. Hence, the primary-primary configuration will vanish when Wick's theorem is applied in the same loop momentum vectors as those that appear on the arguments of the second-order kernel. 
 
 We have already computed the integrals over $\veck_1,\;\veck_2,\; {\rm and}\;\veck_3$ in $\S$ \ref{sec:S-S_conf}, given by Eq. (\ref{eq:I_SS_k3}). We write them out explicitly again here since they will have different angular momenta dependence from those of Eq. (\ref{eq:I_SS_k3}):
\begin{align}\label{eq:I_PP_k2}
&I^{(k_1)}(\vecr_1,-\vecs,-\vecr'_1) =(4\pi)^2 \sum_{L_{11},L'_{1s},L'_{11}} C_{L_{11},L'_{1s},L'_{11}} \Upsilon_{L_{11},L'_{1s},L'_{11}} \mathcal{G}_{L_{11},L'_{1s},L'_{11}}  \nonumber \\
& \qquad \qquad \qquad \qquad \qquad  \times g^{[0]}_{L_{11},L'_{1s},L'_{11}}(r_1,s,r'_1) \mathcal{P}_{L_{11},L'_{1s},L'_{11}}(\hatr_1,-\hats,-\hatr'_1),    
\end{align}
\begin{align}\label{eq:I_PP_k2}
&I^{(k_2)}(\vecr_2,-\vecs,-\vecr'_2) =(4\pi)^2 \sum_{L_{22},L'_{2s},L'_{22}} C_{L_{22},L'_{2s},L'_{22}} \Upsilon_{L_{22},L'_{2s},L'_{22}} \mathcal{G}_{L_{22},L'_{2s},L'_{22}}  \nonumber \\
& \qquad \qquad \qquad \qquad \qquad  \times g^{[0]}_{L_{22},L'_{2s},L'_{22}}(r_2,s,r'_2) \mathcal{P}_{L_{22},L'_{2s},L'_{22}}(\hatr_2,-\hats,-\hatr'_2),    
\end{align} 
\begin{align}\label{eq:I_PP_k3}
&I^{(k_3)}(\vecr_3,-\vecs,-\vecr'_3) =(4\pi)^2 \sum_{L_{33},L'_{3s},L'_{33}} C_{L_{33},L'_{3s},L'_{33}} \Upsilon_{L_{33},L'_{3s},L'_{33}} \mathcal{G}_{L_{33},L'_{3s},L'_{33}}  \nonumber \\
& \qquad \qquad \qquad \qquad \qquad  \times g^{[0]}_{L_{33},L'_{3s},L'_{33}}(r_3,s,r'_1) \PP_{L_{33},L'_{3s},L'_{33}}(\hatr_3,-\hats,-\hatr'_3).    
\end{align}
\begin{figure}[h]
\centering
\includegraphics[scale=0.5]{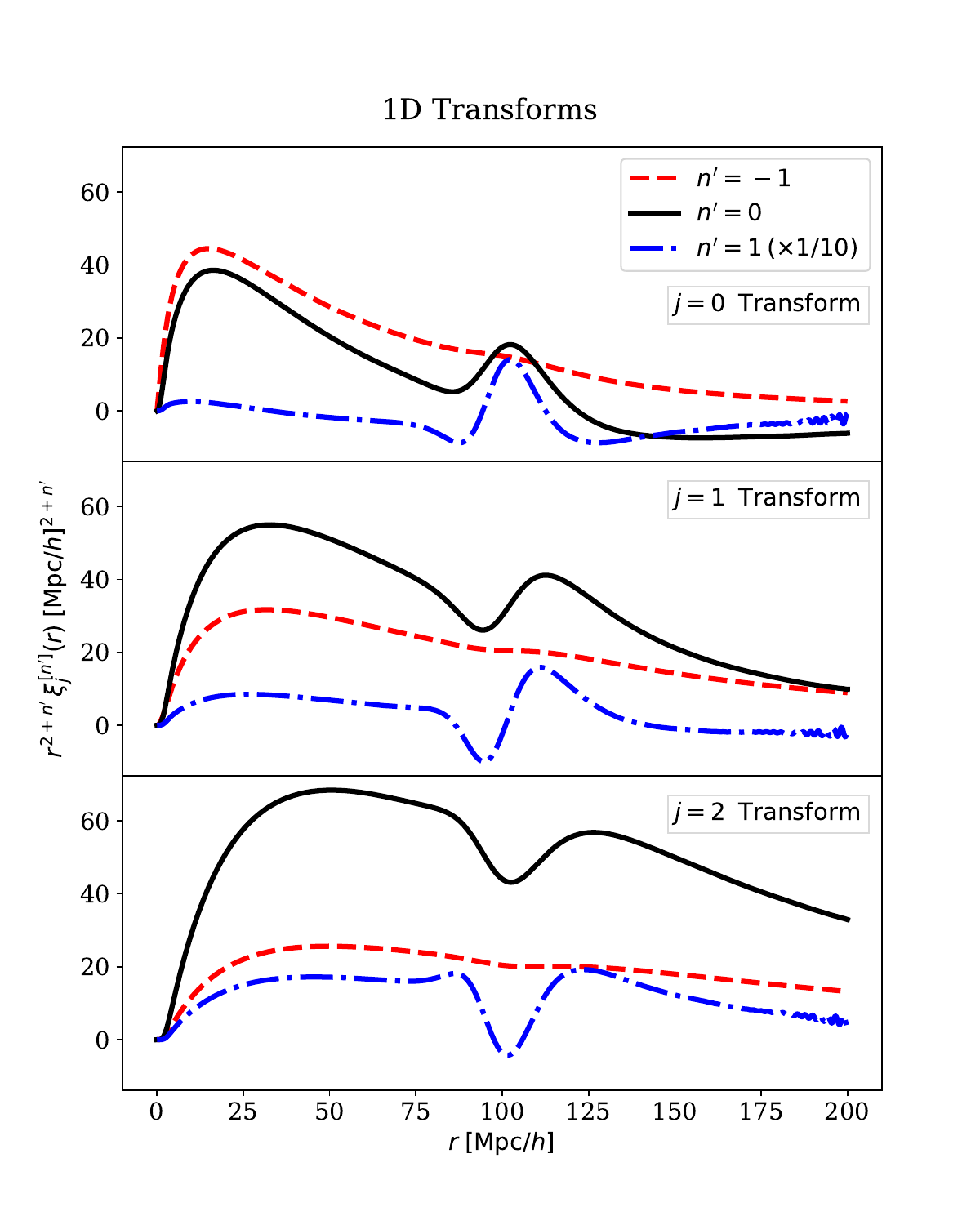}
\caption{1D plot of the integral Eq. (\ref{eq:xi_def}) evaluated at different values of the power $n'$ and spherical Bessel function order $j$. The plot and the figure caption have been taken from section 3 of \cite{Ortola_4PCF}. Each panel shows the integral for a given $j$, and within each panel we show different powers $n'$. This integral falls off as $1/r^{2+n'}$, as we shown in \cite{Ortola_4PCF}; therefore we have weighted the integral by $r^{2+n'}$ to take out its fall-off. In the \textit{top panel}, the black solid line for $n'=0$, $j=0$ integral reproduces the 2-Point Correlation Function (2PCF) as expected. In the \textit{lower panel}, the black solid line for the $n'=0\;j=2$ reproduces the quadrupole of the 2PCF but with a sign flip. The sign flip occurs because, when taking the inverse Fourier transform of the power spectrum to find the 2PCF's  quadrupole, the integration picks up a $(-i)^{\ell}$ from the plane-wave expansion, which is later evaluated at $\ell=2$. We also find that the curves with $n'=0,1$ in the \textit{middle panel} correspond to the derivatives of the curves with $n'=-1,0$ in the \textit{top panel}, as shown in \cite{Ortola_4PCF}. It is also notable that the $n'=-1$ curves are smoother than the curves with $n'=0,1$. The reason for this is that $n'=-1$ brings a factor of $1/k$ in the integrand, which serves as a broad low-pass filter, smoothing the curves. Meanwhile, the $n'=0,1$ curves have factors of $k$ that serve as high-pass filters, bringing out sharper features in the larger scales. The blue dotted line has been re-scaled by a factor of $1/10$ in all three panels.}
\label{fig:1-d_int}
\end{figure}
\clearpage
Evaluating the Wick contraction between $\vecq_1$ and $\vecq_2$, and performing the integral over $\veck'_0$, we obtain:
\begin{align}
&{\rm Cov}_{4, \rm 1L}^{[3],{\rm P-P}}(\mathbf{R},\mathbf{R}') = \int_{\vecs}I^{(k_1)}(\vecr_1,-\vecs,-\vecr'_1)I^{(k_2)}(\vecr_2,-\vecs,-\vecr'_2)I^{(k_3)}(\vecr_3,-\vecs,-\vecr'_3)   \nonumber\\
& \qquad \qquad \qquad \quad \times \int_{\veck_0,\vecq_2}e^{-i\veck_0\cdot(\vecr_0-\vecs-\vecr'_0)} \W(-\vecq_2,\vecq_2,\veck_0) P(q_2) P(k_0).
\end{align}
We proceed by expanding the third-order kernel using Eq. (\ref{eq:W_3_def}), and solve the $\vecq_2$ integral:
\begin{align}
&I^{(q_2)}(r) = (-1)^{n_{12}+n_{13}} \int_{\vecq_2}  q_2^{n_1'+n'_2+n} \;Y_{j_{12},m_{j_{12}}}(\hatq_2)Y_{j_{13},m_{j_{13}}}(\hatq_2)\nonumber \\
& \qquad \qquad \qquad\qquad \quad \times Y_{j,m_j}(\hatq_2)Y_{\ell_2,m_2}(\hatq_2)\;j_{\ell_2}(q_2 r) P(q_2)\nonumber \\
& \qquad \qquad \qquad\quad\;\;\;= 2\pi^2\;(-1)^{n_{12}+n_{13}}\; \sum_{\ell,m}G_{m_{j_{12}},m_j,-m}^{j_{12},j,\ell}G_{m_{j_{13}},m,m_2}^{j_{13},\ell,\ell_2}\; \xi_{\ell_2}^{[n_1'+n'_2+n]}(r),
\end{align}
where $G$ is the Gaunt coefficient as defined in Eq. (\ref{eq:Gaunt_Coeff}), and the radial integral we have defined as:
\begin{align}\label{eq:xi_def}
\xi_{j}^{[n]}(r) \equiv \int\frac{dk \;k^{n+2}}{2\pi^2}\;j_{j}(kr)P(k).
\end{align}
We remind the reader the sums from the $W^{(3)}$ kernel are omitted above since they are used across multiple equations, but we introduce them back in Eq. (\ref{eq:PP_res}).

\begin{figure}[h!]
\centering
\includegraphics[scale=0.6]{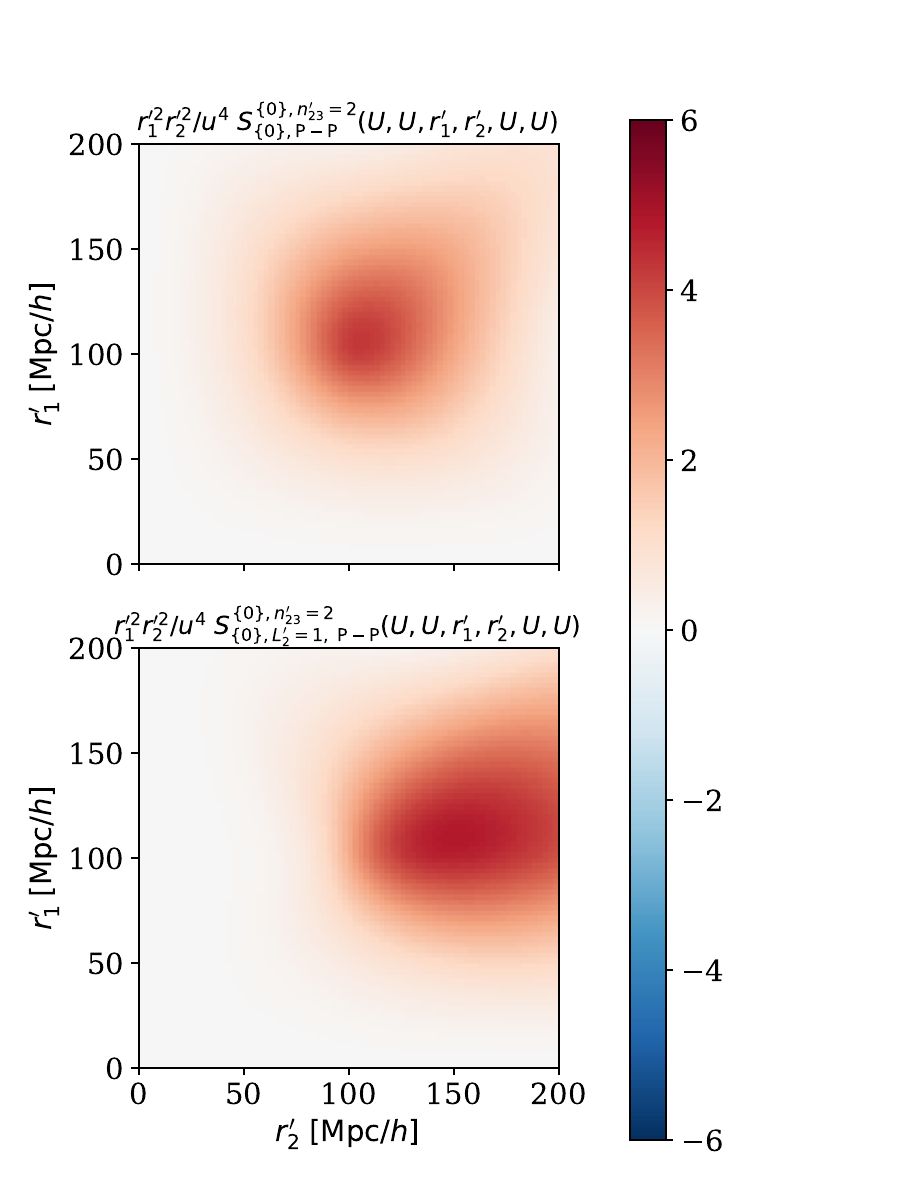}
\caption{Here, we show Eq. (\ref{eq:Sint_PP}) with fixed arguments at $ U \equiv 100 \;\left[{\rm Mpc}/h\right]$, except for $r'_1$ and $r'_2$. We also chose to evaluate $n' = n=0$, $n'_{23}=2$ and all angular momenta $\{L\} = 0$, except for $L'_2 = \left\{0,1\right\}$. The \textit{upper panel} shows the integral for $L'_2=0$, while the \textit{lower panel} shows the integral for $L'_2=1$. Since the 4PCF can be approximated as the square of the 2PCF on large scales, $(\xi_{0})^{2}(r) \sim (1/r^2)^2$, we have weighted the integral by $r_1^{'2}r_2^{'2}/u^{4}$, with $u \equiv 10 \;\left[{\rm Mpc}/h\right]$ to take out its fall-off. The $S$ integral is built from the intermediate radial integrals $\xi$, $g$, and $h$. The $\xi$ integral exhibits a bump around $U$ for $n=0,\;j=0$, as expected since this corresponds to the two-point correlation function (see Fig. \ref{fig:1-d_int}). The $g$ integral produces a rectangular boundary (Fig. \ref{fig:gint2}), while the $h$ integral generates a blob near the chosen value of $U$ (Fig. \ref{fig:hint}). When combined, the integration of the rectangular boundaries and the blob yields the oval-shaped regions visible in both panels, with the bump in $\xi$ ensuring that most of the behavior remains centered around $U$.}
\label{fig:S_PP_int}
\end{figure}

With the already expanded third-order kernel, we proceed by evaluating the $\veck_0$ integral:
\begin{align}
&I^{(k_0)}(\vecr_0,-\vecs,-\vecr'_0) = \int_{\veck_0}  k_0^{n_3'-n} e^{-i\veck_0\cdot(\vecr_0-\vecs-\vecr'_0)}\;Y_{j_{12},m_{j_{12}}}(\hatk_0) \nonumber \\
& \qquad \qquad \qquad \qquad  \times Y_{j_{13},m_{j_{13}}}(\hatk_0)Y_{j,m_j}(\hatk_0)Y_{\ell_2,m_2}(\hatk_0)\;j_{\ell_2}(k_0 r) P(k_0)\nonumber \\
&\qquad \qquad\qquad \quad \;\;= (4\pi)^2 \sum_{L_{00},L_{0s},L'_{00}} C_{L_{00},L_{0s},L'_{00}} \Upsilon_{L_{00},L_{0s},L'_{00}} \mathcal{L}_{L_{00},L_{0s},L'_{00}}^{j_{12},j_{13},j,\ell_2,\{m\}}  \nonumber \\
& \qquad \qquad  \qquad \qquad\times h^{[n'_3-n]}_{L_{33},L'_{3s},L'_{33},\ell_2}(r_0,s,r'_0,r) \PP_{L_{00},L_{0s},L'_{00}}(\hatr_0,-\hats,-\hatr'_0), 
\end{align}
where we have expanded the complex exponential in terms of the isotropic basis functions using Eq. (\ref{eq:PWE}), and then evaluated the angular integral using Eq. (\ref{eq:Fancy_L_def}). With the above result, we only lack to evaluate the $\hats$ integral, which we obtain in $\S$\ref{sec:s_hat_PP_Evaluation}. Therefore, the Primary-Primary configuration results in:
\begin{empheq}[box=\widefbox]{align}\label{eq:PP_res}
&{\rm Cov}_{4, \rm 1L}^{[3],{\rm P-P}} (\mathbf{R},\mathbf{R}') = (2\pi)^{15} \;(4\pi)^{6}\;2\pi^2 \sum_{\rm All}\nonumber \\
& \qquad \qquad \times (-1)^{L_{00} + L_{11} + L_{22} + L_{33}+n_{12}+n_{13}} \nonumber \\
& \qquad \qquad \times \overline{\mathcal{Q}}_{\rm (P-P)}^{\Lambda,\Lambda_s,\Lambda'}\mathcal{S}^{P}_{\Lambda_s}\mathcal{C}^{\Lambda_s}_{\mathbf{0}} d_{j_{12}}^{(n_{12})}d_{j_{13}}^{(n_{13})}d_{j_{23}}^{(n_{23})} D_j \;\mathcal{D}_{n'_1,n'_2,n'_3,n'_{23}}^{n_{12},n_{13},n_{23}} \nonumber \\
& \qquad \qquad \times C_{L_{00},L_{0s},L'_{00}} \Upsilon_{L_{00},L_{0s},L'_{00}} \mathcal{L}_{L_{00},L_{0s},L'_{00}}^{j_{12},j_{13},j,\ell_2,\{m\}} \nonumber \\
& \qquad \qquad \times C_{L_{11},L'_{1s},L'_{11}} \Upsilon_{L_{11},L'_{1s},L'_{11}} \mathcal{G}_{L_{11},L'_{1s},L'_{11}}\nonumber \\
& \qquad \qquad \times  C_{L_{22},L'_{2s},L'_{22}} \Upsilon_{L_{22},L'_{2s},L'_{22}} \mathcal{G}_{L_{22},L'_{2s},L'_{22}} \nonumber \\
& \qquad \qquad \times C_{L_{33},L'_{3s},L'_{33}} \Upsilon_{L_{33},L'_{3s},L'_{33}} \mathcal{G}_{L_{33},L'_{3s},L'_{33}} \nonumber \\
& \qquad \qquad \times G_{m_{j_{12}},m_j,-m}^{j_{12},j,\ell}G_{m_{j_{13}},m,m_2}^{j_{13},\ell,\ell_2}\nonumber \\
& \qquad \qquad \times S_{\{L\}, {\rm P-P}}^{(\{L_{is,r}\})}(r_0,r'_0,r_1,r'_1,r_2,r'_2,r_3,r'_3) \nonumber \\
& \qquad \qquad \times\PP_{\Lambda}(\hatR)\PP_{\Lambda'}(\hatR')+\;71\;{\rm perms.}
\end{empheq}
In the sum above, ``All'' denotes the sum over all angular momentum and magnetic quantum numbers appearing in the expression. The summation range for each angular momentum variable is defined in the equation where that variable is first introduced. For convenience, Table~\ref{table:1} lists all angular momentum coupling coefficients together with the equations in which they are defined. We have defined $S$ as:
\begin{align}\label{eq:Sint_PP}
&S_{\{L\}, {\rm P-P}}^{(\{L_{is,r}\})}(r_0,r'_0,r_1,r'_1,r_2,r'_2,r_3,r'_3) \nonumber \\
& \qquad  \equiv \int dr \;r^{n_{23}-1}\;\xi_{\ell_2}(r)\int ds\;s^2\; g_{L_{11},L_{1s},L'_{11}}^{[0]} (r_1,s,r'_1) \;g_{L_{22},L_{2s},L'_{22}}^{[0]} (r_2,s,r'_2)\nonumber \\
& \qquad  \times g_{L_{33},L_{3s},L'_{33}}^{[0]} (r_3,s,r'_3) \; h^{[n'_3-n]}_{L_{33},L'_{3s},L'_{33},\ell_2}(r_0,s,r'_0,r).
\end{align} 
We use the subscript $\{L\}$ to denote all the angular momentum indices that affect one of the position-space variables (\textit{i.e.}, the $r_0$, $r'_0$,$\cdots$ variables). The superscript $(\{L_{is,r}\})$ denotes the set of all angular momentum indices that are coupled to $s$ and $r$ in the spherical Bessel functions.  Numerical evaluation of this radial integral for the lowest order angular momenta ($L=\{0,1\}$) is shown in Fig. \ref{fig:S_PP_int}. The behavior of this radial integral is smooth and predominantly positive over the plotted domain. In the upper panel, it is localized around \(r_1' \sim r_2' \sim U\), indicating sensitivity to matching between separations and the fixed external scale \(U=100\,\mathrm{Mpc}/h\). The lower panel is broader, but again is suppressed for \(r_1',r_2' \ll U\) and becomes appreciable once the primed separations are comparable to \(U\). Its net effect on the full NLO 4PCF covariance must nevertheless be assessed only after including the remaining structures, angular couplings, and permutations.

\section{Discussion and Conclusion}\label{sec:Conclussion}
In this work, we have developed a systematic framework to compute the covariance of the 4PCF beyond the Gaussian approximation using the third-order density contrast and galaxy biasing. The central step was to unify the relevant third-order perturbative contributions into a single effective kernel, $W^{(3)}$, defined in Eq.~(\ref{eq:W_3_def}). This formulation makes it possible to treat the full class of third-order contributions in a common language, without tying the derivation to any one specific kernel choice.

Applying Wick’s theorem to the density contrasts, we showed that the loop contributions can be organized into only three distinct classes, according to how the loop is embedded within the tetrahedral configuration: secondary-secondary, primary-secondary, and primary-primary. The explicit expressions for these contributions are given in Eqs.~(\ref{eq:SS_res}), (\ref{eq:PS_res}), and (\ref{eq:PP_res}), and the corresponding configurations are illustrated in Fig.~\ref{fig:dc_configurations}. Thus, despite the apparent complexity of the one-loop calculation, the full third-order contribution to the connected 4PCF covariance can be reduced to three universal building blocks.

The practical implication is that the covariance sourced by third-order densities can be constructed from these three structures alone, with the detailed perturbative physics entering only through the generalized kernel $W^{(3)}$. This provides a direct and systematic route to building the full third-order contribution to the 4PCF covariance. The complementary contribution sourced by second-order densities is derived in our companion paper~\cite{Ortola_4PCF_CovI}, and together these results establish the full NLO 4PCF covariance framework.

Relative to the Gaussian approximation, in which Wick’s theorem reduces the covariance to products of two-point functions, the terms computed here capture the one-loop corrections sourced by third-order densities. These represent part of the full NLO covariance and are expected to become increasingly relevant on mildly nonlinear scales, where the Gaussian approximation is no longer sufficient. In Appendix~\ref{sec:Correction_Scale}, we discuss the scales on which these corrections are expected to become important and compare their contribution to the Gaussian term.

At the same time, the present calculation has a number of limitations. To keep the derivation tractable, we have not included redshift-space distortions, survey window effects, or shot noise. In addition, our treatment is based on standard perturbation theory and does not incorporate ultraviolet counterterms or infrared resummation. The results presented here should therefore be viewed as a first step toward a more complete and survey-realistic model for the 4PCF covariance.

Since a formal result of this kind is only useful if it can be implemented in practice, we have also examined the efficiency and numerical behavior of the resulting expressions. In particular, we study the computational cost of the formalism in Appendix~\ref{sec:Computational_Cost} and the convergence of the angular momentum sums in Appendix~\ref{sec:Convergence_Test}. These results show that the analytical reformulation developed here provides a substantial simplification over a brute-force evaluation of the loop integrals.

Looking ahead, several important extensions remain. A key next step will be to combine the present third-order results with the second-order contributions into a single complete NLO covariance implementation. It will also be important to extend the framework to include redshift-space distortions and survey geometry, and to validate the resulting predictions against $N$-body simulations and mock catalogs. These developments are necessary to assess the range of validity of the perturbative treatment and to produce survey-ready analytical covariance matrices for DESI and future galaxy surveys.

We have also made available a public \href{https://doi.org/10.5281/zenodo.20528953}{Zenodo repository} containing code to reproduce selected figures from this paper.

\acknowledgments
We thank the Slepian research group for useful discussions and comments on this work. ZS especially thanks Bob Cahn for useful discussions. This publication was made possible through the support of Grant 63041 from the John Templeton Foundation. The opinions expressed in this publication are those of the author(s) and do not necessarily reflect the views of the John Templeton Foundation. ZS acknowledges funding from NASA grant number 80NSSC24M0021; ZS acknowledges funding from UF Research AI award \#00133699. The authors acknowledge UFIT Research Computing for providing computational resources on HiPerGator and support that have contributed to the research results reported in this publication.

\appendix

\section{Generalizing the Third-Order Density Contrasts for Use in the Covariance Matrix}\label{sec:Generalization_delta_3}
In this section we generalize the $\delta^{(3)},\;\delta^{3}_{\rm lin.},\; \delta_{\rm lin.}\mathcal{G}^{(2)}$, $\mathcal{G}^{(3)}$ and $\Gamma^{(3)}$ to a single third-order density contrast in Fourier space. We do so using a generalized third-order kernel we term $W^{(3)}$, defined in Eq. (\ref{eq:W_3_def}). 

\subsection{Third-Order Density Kernel}
We begin by expressing the third-order density contrast in terms of its inverse FT:
\begin{align}\label{eq:delta_3_def_FS}
&\delta^{(3)}(\vecx + \vecr_i) = \int \frac{d^{3}\veck_i}{(2\pi)^3} \;e^{-i \veck_i\cdot (\vecx+\vecr_i)} \widetilde{\delta}^{(3)}(\veck_i) \nonumber \\ 
& = \int \frac{d^{3}\veck_i}{(2\pi)^3} \;e^{-i \veck_i\cdot (\vecx+\vecr_i)} \int d^3\vecq_1 \;d^3\vecq_2\;d^3\vecq_3\; \dD(\veck_i - \sum_{l=1}^{3}\vecq_l)\nonumber \\ 
& \qquad\times F^{(3)}(\vecq_1,\vecq_2,\vecq_3) \;\widetilde{\delta}_{\rm lin.}(\vecq_1) \widetilde{\delta}_{\rm lin.}(\vecq_2) \widetilde{\delta}_{\rm lin.}(\vecq_3)\nonumber\\
& = \int d^3\vecq_1 \;d^3\vecq_2\;d^3\vecq_3 \;e^{-i (\vecq_1+\vecq_2+\vecq_3)\cdot (\vecx+\vecr_i)} F^{(3)}(\vecq_1,\vecq_2,\vecq_3) \;\widetilde{\delta}_{\rm lin.}(\vecq_1) \widetilde{\delta}_{\rm lin.}(\vecq_2)\widetilde{\delta}_{\rm lin.}(\vecq_3).
\end{align}
To obtain the second equality, we used the definition of the third-order density contrast in terms of the symmetrized third-order kernel $F^{(3)}$ \cite{Bernardeau}. To obtain the third equality, we evaluated the Dirac delta function to perform the $\veck_i$ integral. The $F^{(3)}$ kernel is defined as \cite{Farshad, Ortola_4PCF}:
\begin{align}\label{F3_def}
 & F^{(3)}\left( \mathbf{k}_1,\mathbf{k}_2,\mathbf{k}_3\right) = \frac{2k_{123}^{2}}{54} \left[ \frac{\mathcal{L}_{1}(\mathbf{\widehat{k}}_1\cdot \mathbf{\widehat{k}}_{23})}{k_1 k_{23}}G^{(2)}\left( \mathbf{k}_2,\mathbf{k}_3\right) + 2 \;{\rm perm.} \right]  \nonumber \\ 
& \qquad   + \frac{7}{54} \left[ \frac{1}{k_{23}} \left\{ k_1 \mathcal{L}_1 (\mathbf{\widehat{k}}_1\cdot \mathbf{\widehat{k}}_{23}) + k_2 \mathcal{L}_1 (\mathbf{\widehat{k}}_2\cdot \mathbf{\widehat{k}}_{23}) + k_3 \mathcal{L}_1 (\mathbf{\widehat{k}}_3\cdot \mathbf{\widehat{k}}_{23}) \right\} G^{(2)}\left( \mathbf{k}_2,\mathbf{k}_3\right) + 2 \; {\rm perm.}\right]  \nonumber \\ 
& \qquad+ \frac{7}{54} \left[ \frac{1}{k_{1}} \left\{ k_1  + k_2 \mathcal{L}_1 (\mathbf{\widehat{k}}_2\cdot \mathbf{\widehat{k}}_1) + k_3 \mathcal{L}_1 (\mathbf{\widehat{k}}_3\cdot \mathbf{\widehat{k}}_1) \right\} F^{(2)}\left( \mathbf{k}_2,\mathbf{k}_3\right) + 2 \; {\rm perm.}\right],
\end{align}
where the $F^{(2)}$ and $G^{(2)}$ kernels are given by:
\begin{align}\label{eq:F2_def}
&F^{(2)}\left(\mathbf{k}_2,\mathbf{k}_3\right) = \frac{17}{21} \mathcal{L}_{0}(\mathbf{\widehat{k}}_2 \cdot \mathbf{\widehat{k}}_3) + \frac{1}{2} \left(\frac{k_2}{k_3}+\frac{k_3}{k_2}\right)\mathcal{L}_{1}(\mathbf{\widehat{k}}_2 \cdot \mathbf{\widehat{k}}_3) + \frac{4}{21}\mathcal{L}_{2}(\mathbf{\widehat{k}}_2 \cdot \mathbf{\widehat{k}}_3)\nonumber \\
&\qquad \qquad \quad\;\;=4\pi\sum_{j=0}^2\sum_{n=-1}^1 c_{j,n}^{(F)} \;\mathcal{P}_{j}(\hatk_2,\hatk_3)\; k_2^n k_3^{-n},
\end{align}
\begin{align}\label{eq:G2_def}
&G^{(2)}\left(\mathbf{k}_2,\mathbf{k}_3\right) = \frac{13}{21} \mathcal{L}_{0}(\mathbf{\widehat{k}}_2 \cdot \mathbf{\widehat{k}}_3) + \frac{1}{2} \left(\frac{k_3}{k_2}+\frac{k_2}{k_3}\right)\mathcal{L}_{1}(\mathbf{\widehat{k}}_2 \cdot \mathbf{\widehat{k}}_3) + \frac{8}{21}\mathcal{L}_{2}(\mathbf{\widehat{k}}_2 \cdot \mathbf{\widehat{k}}_3)\nonumber \\
&\qquad \qquad \quad\;\;=4\pi\sum_{j=0}^2\sum_{n=-1}^1 c_{j,n}^{(G)} \;\mathcal{P}_{j}(\hatk_2,\hatk_3)\; k_2^n k_3^{-n},
\end{align}
where we define $\hatk_{23}\equiv (\veck_2+\veck_3)/|\veck_{23}|$ with $\veck_{23} \equiv \veck_2 +\veck_{3}$. The values of $c^{(F)}_{j,n}$ are $c^{(F)}_{0,0}=17/21$, $c^{(F)}_{1,1}=c^{(F)}_{1,-1}=-\sqrt{3}/2$, and $c^{(F)}_{2,0} = 4\sqrt{5}/21$; the rest of constants are zero. The values of $c^{(G)}_{j,n}$ are $c^{(G)}_{0,0}=13/21$, $c^{(G)}_{1,1}=c^{(G)}_{1,-1}=-\sqrt{3}/2$, and $c^{(G)}_{2,0} = 8\sqrt{5}/21$; the rest of constants are zero. The $\mathcal{L}_{\ell}$ refers to the Legendre polynomial of order $\ell$ and these second-order kernels can be written in terms of the isotropic basis functions as already derived in \cite{Ortola_4PCF}:
\begin{align}\label{eq:W2_def}
\nonumber \\
W^{(2)}_{\mu}(\veck_2,\veck_3)=4\pi\sum_{j, n} c_{j,n}^{(\mu)} \;\mathcal{P}_{j}(\hatk_2,\hatk_3) k_2^n k_3^{-n},
\end{align}
where $\mu$ runs over $\{F,G\}$ and indicates if this is an $F^{(2)}$ or $G^{(2)}$ kernel. 

In the main text, we need to write the isotropic basis function in the equation above in terms of the spherical harmonics to evaluate each angular dependence separately. We express the 2-argument isotropic basis functions in terms of the spherical harmonics as \cite{Cahn_Iso}:
\begin{align}\label{eq:2_arg_Iso_to_Sph_har}
\PP_j(\hatk_2,\hatk_3) = D_{j}\sum_{m_j=-j}^j Y_{j,m_j}(\hatk_2)Y^*_{j,m_j}(\hatk_3),
\end{align}
where we have defined:
\begin{align}\label{eq:D_j_def}
D_j\equiv(-1)^j/s_{j}^{(\rm I)},
\end{align}
with $s_{j}^{(\rm I)}$ defined in Eq. (\ref{eq:def_I_k1}). 

Hence, the third-order kernel can be written as \cite{Ortola_4PCF}:
\begin{align}\label{eq:W3_expression}
&W^{(3)}\left( \mathbf{k}_1,\mathbf{k}_2,\mathbf{k}_3\right) = \sum_{n'_1,n'_2,n'_3=0}^{4}\; \sum_{n_{12},n_{13},n_{23} = 0}^{4} \;\sum_{n'_{23}=0,2}\;\sum_{\mu=\{F,G\}}\;\mathcal{D}_{n'_1,n'_2,n'_3,n'_{23}}^{n_{12},n_{13},n_{23}}\frac{k_{1}^{n'_1}k_{2}^{n'_2}k_{3}^{n'_3}}{k_{23}^{n'_{23}}} \nonumber \\
& \qquad \qquad \qquad \quad \times (\mathbf{\widehat{k}}_1 \cdot \mathbf{\widehat{k}}_2)^{n_{12}}(\mathbf{\widehat{k}}_1 \cdot \mathbf{\widehat{k}}_3)^{n_{13}} (\mathbf{\widehat{k}}_2 \cdot \mathbf{\widehat{k}}_3)^{n_{23}} W^{(2)}_{\mu}\left(\mathbf{k}_2,\mathbf{k}_3\right), 
\end{align}
with $\mathcal{D}$ ensuring only the right combinations of exponents $n$ and $n'$ result in a non-zero $W^{(3)}$. We do not explicitly state all the non-zero values of $\mathcal{D}$, but we show an example of how to obtain those values. The first term in Eq. (\ref{F3_def}) can be evaluated as follows:
\begin{align}\label{eq:example_Cal_D_1}
&\frac{2 k_{123}^2}{54}\frac{\mathcal{L}_{1}(\hatk_1\cdot\hatk_{23})}{k_1k_{23}} = \frac{2}{54\;k_1k_{23}^2}\bigg[k_1^2+2k_1k_2\left(\hatk_1\cdot\hatk_2\right)+k_2^2+2k_1k_3\left(\hatk_1\cdot\hatk_3\right)\nonumber \\
& \qquad \qquad\qquad \qquad \qquad  +k_3^2+2k_2k_3\left(\hatk_2\cdot\hatk_3\right)\bigg]\bigg[\left(\hatk_1\cdot\hatk_2\right)+ \left(\hatk_1\cdot\hatk_3\right)\bigg],
\end{align}
where we have expanded the $k_{123}^2$ term explicitly and also $\mathcal{L}_{1}(\hatk_1\cdot\hatk_{23})$. We have excluded the $G^{(2)}$ kernel above, but it is implicitly included in the example. After distributing the terms within the square brackets, we can obtain the value for each combination of the $\mathcal{D}$ constant:
\begin{align}\label{eq:example_Cal_D_2}
&\frac{2\; k_1\left(\hatk_1\cdot\hatk_2\right)}{54\; k_{23}^2} \to \mathcal{D}^{1,0,0}_{1,0,0,2} = \frac{2}{54},\nonumber\\
&\frac{4\; k_2\left(\hatk_1\cdot\hatk_2\right)^2}{54\; k_{23}^2} \to \mathcal{D}^{2,0,0}_{0,1,0,2} = \frac{4}{54},\; {\rm etc.}
\end{align}

\subsection{Cubic Linear Density Contrast}
We will now show that $\widetilde{\delta}_{\rm lin.}^3$ can be expressed in the same form but with different coefficients:
\begin{align}\label{eq:delta_1_sq_def}
&\delta_{\rm lin.}^3(\vecx + \vecr_i) = \delta_{\rm lin.}(\vecx + \vecr_i)\delta_{\rm lin.}(\vecx + \vecr_i) \delta_{\rm lin.}(\vecx + \vecr_i) \nonumber \\
& = \int \frac{d^{3}\veck_i}{(2\pi)^3}\; e^{-i \veck_i\cdot (\vecx+\vecr_i)} \widetilde{\delta}_{\rm lin.}(\veck_i)  \int \frac{d^{3}\veck''_i}{(2\pi)^3} \;e^{-i \veck''_i\cdot (\vecx+\vecr_i)}\widetilde{\delta}_{\rm lin.}(\veck''_i) \nonumber \\
& \quad \times \int \frac{d^{3}\veck^{'''}_i}{(2\pi)^3}\; e^{-i \veck^{'''}_i\cdot (\vecx+\vecr_i)} \widetilde{\delta}_{\rm lin.}(\veck^{'''}_i) \nonumber \\
&= \int \frac{d^{3}\veck_i\;d^{3}\veck''_i\;d^{3}\veck^{'''}_i}{(2\pi)^9}\; e^{-i (\veck_i+\veck''_i+\veck^{'''}_i)\cdot (\vecx+\vecr_i)} \widetilde{\delta}_{\rm lin.}(\veck_i) \widetilde{\delta}_{\rm lin.}(\veck''_i)\widetilde{\delta}_{\rm lin.}(\veck^{'''}_i) \nonumber\\
& = (2\pi)^{9} \lim_{F^{(3)}\to 1} \delta^{(3)}(\vecx + \vecr_0).
\end{align}
In the third equality, we observed that the cubic linear density is simply what would result if we took the limit that $F^{(3)}\to 1$ in our usual computation of the second-order density by comparing with Eq. (\ref{eq:delta_3_def_FS}).

\subsection{Density Contrast with Second-Order Galilean Kernel}
Next we show that $\delta_{\rm lin.}\mathcal{G}^{(2)}$ can be expressed in the same format as the third-order kernel but with different coefficients:
\begin{align}\label{eq:delta_mathcalG2}
&\delta_{\rm lin.}(\vecx + \vecr_i)\mathcal{G}^{(2)}(\vecx + \vecr_i)  \nonumber \\
& = \int \frac{d^{3}\veck_i}{(2\pi)^3}\; e^{-i \veck_i\cdot (\vecx+\vecr_i)} \widetilde{\delta}_{\rm lin.}(\veck_i)  \int \frac{d^{3}\veck'_i}{(2\pi)^3} \;e^{-i \veck'_i\cdot (\vecx+\vecr_i)}\int d^3\vecq_1 \;d^3\vecq_2\; \dD(\veck'_i - \vecq_1 - \vecq_2)\nonumber \\ 
& \qquad \quad\times \mathcal{G}^{(2)}(\vecq_1,\vecq_2) \;\widetilde{\delta}_{\rm lin.}(\vecq_1) \widetilde{\delta}_{\rm lin.}(\vecq_2) \nonumber \\
& = \int \frac{d^{3}\veck_i\;d^{3}\vecq_1\;d^{3}\vecq_2}{(2\pi)^6}\; e^{-i (\veck_i+\vecq_1+\vecq_2)\cdot (\vecx+\vecr_i)} \;\mathcal{G}^{(2)}(\vecq_1,\vecq_2)\;\widetilde{\delta}_{\rm lin.}(\veck_i) \widetilde{\delta}_{\rm lin.}(\vecq_1)\widetilde{\delta}_{\rm lin.}(\vecq_2) \nonumber\\
& = (2\pi)^{3} \lim_{F^{(3)}\to (2/3)[\mathcal{L}_2(\hatq_1\cdot\hatq_2)-2/3]} \delta^{(3)}(\vecx + \vecr_0),
\end{align}
where the second-order Galilean kernel in Fourier space is defined as:
\begin{align}
\mathcal{G}^{(2)}(\vecq_1,\vecq_2) \equiv \frac{2}{3}\left[\mathcal{L}_2(\hatq_1\cdot\hatq_2)-\frac{2}{3}\right].
\end{align}
In the third equality, we conclude that the linear density with the second-order Galilean kernel is what would result if we took the limit that $F^{(3)}\to (2/3)[\mathcal{L}_2(\hatq_1\cdot\hatq_2)-2/3]$ in our usual computation of the third-order density.

\subsection{Third-Order Galilean Kernel}
We continue with the third-order Galilean kernel in Fourier space:
\begin{align}
\mathcal{G}^{(3)}(\veck_1,\veck_2,\veck_3) = \sum_{n_{12},n_{13},n_{23}=0}^{2} g_{n_{12},n_{13},n_{23}} (\hatk_1\cdot\hatk_2)^{n_{12}} (\hatk_1\cdot\hatk_3)^{n_{12}} (\hatk_2\cdot\hatk_3)^{n_{23}},
\end{align}
where $g_{1,1,1}=2$, $g_{2,0,0}=g_{0,2,0}=g_{0,0,2}=-1$; all other combinations are zero. Therefore, we obtain:
\begin{align}\label{eq:G3_def}
&\mathcal{G}^{(3)}(\vecx + \vecr_i) = \int \frac{d^{3}\veck_i}{(2\pi)^3} \;e^{-i \veck_i\cdot (\vecx+\vecr_i)}\int d^3\vecq_1 \;d^3\vecq_2\;d^3\vecq_3\; \dD(\veck_i - \vecq_1 - \vecq_2- \vecq_3)\nonumber \\ 
& \qquad \qquad\qquad\qquad\times \mathcal{G}^{(3)}(\vecq_1,\vecq_2,\vecq_3) \;\widetilde{\delta}_{\rm lin.}(\vecq_1) \widetilde{\delta}_{\rm lin.}(\vecq_2)\widetilde{\delta}_{\rm lin.}(\vecq_3) \nonumber \\
& \qquad \qquad \quad\;\; =  \lim_{W^{(3)}\to \mathcal{G}^{(3)}} \delta^{(3)}(\vecx+\vecr_i), 
\end{align}
where the limit holds when the second-order kernel on the right hand side of Eq. (\ref{eq:W3_expression}) goes to unity and the magnitudes of the wave vectors are raised to the zeroth power:
\begin{align}
&\lim_{W^{(2)}\to1} \;\lim_{n'_1,n'_2,n'_3,n'_{23}\to 0}\mathcal{D}_{n'_1,n'_2,n'_3,n'_{23}}^{n_{12},n_{13},n_{23}}\frac{k_{1}^{n'_1}k_{2}^{n'_2}k_{3}^{n'_3}}{k_{23}^{n'_{23}}}  (\mathbf{\widehat{k}}_1 \cdot \mathbf{\widehat{k}}_2)^{n_{12}}(\mathbf{\widehat{k}}_1 \cdot \mathbf{\widehat{k}}_3)^{n_{13}} (\mathbf{\widehat{k}}_2 \cdot \mathbf{\widehat{k}}_3)^{n_{23}} W^{(2)}_{\mu}\left(\mathbf{k}_2,\mathbf{k}_3\right)\nonumber \\
& \qquad \qquad \qquad \qquad = g_{n_{12},n_{13},n_{23}} (\hatk_1\cdot\hatk_2)^{n_{12}} (\hatk_1\cdot\hatk_3)^{n_{12}} (\hatk_2\cdot\hatk_3)^{n_{23}}.
\end{align} 

\subsection{Third-Order Gamma Kernel}
We continue with the third-order Gamma kernel in Fourier space as expressed in Eq. (3.44) of \cite{Ortola_4PCF}:
\begin{align}
&\Gamma^{(3)} (\mathbf{k}_1, \mathbf{k}_2, \mathbf{k}_3) \rightarrow \left[ \frac{(\mathbf{\widehat{k}}_1\cdot\mathbf{k}_2)^2 + (\mathbf{\widehat{k}}_1\cdot\mathbf{k}_3)^2 + 2(\mathbf{\widehat{k}}_1\cdot\mathbf{k}_2)(\mathbf{\widehat{k}}_1\cdot\mathbf{k}_3) }{k_{23}^{2}} -1 \right]W^{(2)}_{\mu}(\mathbf{k}_2, \mathbf{k}_3) +\;2\;\text{perms.}\nonumber \\ 
& = \sum_{\rm All} \gamma_{n'_1,n'_2,n'_3,n'_{23}}^{n_{12},n_{13},n_{23}}\frac{k_{1}^{n'_1}k_{2}^{n'_2}k_{3}^{n'_3}}{k_{23}^{n'_{23}}} (\mathbf{\widehat{k}}_1 \cdot \mathbf{\widehat{k}}_2)^{n_{12}}(\mathbf{\widehat{k}}_1 \cdot \mathbf{\widehat{k}}_3)^{n_{13}} (\mathbf{\widehat{k}}_2 \cdot \mathbf{\widehat{k}}_3)^{n_{23}} W^{(2)}_{\mu}\left(\mathbf{k}_2,\mathbf{k}_3\right)+\;2\;\text{perms.}
\end{align}
We observe this kernel can be exactly computed as the third-order kernel without the need for further calculations, as the structure is practically identical. The non-zero values of $\gamma$ are obtained following the same example given in Eqs. (\ref{eq:example_Cal_D_1}) and (\ref{eq:example_Cal_D_2}). 
 
\subsection{Generalized Third-Order Density Contrast}
Below, we present the generalized third-order density contrast and kernel:
\begin{align}\label{eq:delta_i_def}
&\delta_{\beta}^{(3)}(\vecx+\vecr_0) = \int \frac{d^{3}\veck_0}{(2\pi)^3}\; e^{-i \veck_0\cdot (\vecx+\vecr_0)}\int d^3\vecq_1 \;d^3\vecq_2\;d^3\vecq_3\; \dD(\veck_0 - \vecq_1 - \vecq_2-\vecq_3)\nonumber \\ 
& \qquad \qquad\qquad\times W_{\beta}^{(3)}(\vecq_1,\vecq_2,\vecq_3) \;\widetilde{\delta}_{\rm lin.}(\vecq_1) \widetilde{\delta}_{\rm lin.}(\vecq_2)\widetilde{\delta}_{\rm lin.}(\vecq_3), 
\end{align}
with
\begin{align}\label{eq:W_3_def}
&W_{\beta}^{(3)}(\veck_1,\veck_2,\veck_3) = \sum_{\rm All} \mathcal{D}_{n'_1,n'_2,n'_3,n'_{23}}^{n_{12},n_{13},n_{23}}\frac{k_{1}^{n'_1}k_{2}^{n'_2}k_{3}^{n'_3}}{k_{23}^{n'_{23}}} \nonumber \\ 
& \qquad \qquad\qquad\times(\mathbf{\widehat{k}}_1 \cdot \mathbf{\widehat{k}}_2)^{n_{12}}(\mathbf{\widehat{k}}_1 \cdot \mathbf{\widehat{k}}_3)^{n_{13}} (\mathbf{\widehat{k}}_2 \cdot \mathbf{\widehat{k}}_3)^{n_{23}} W^{(2)}_{\beta}\left(\mathbf{k}_2,\mathbf{k}_3\right),
\end{align}
where $\beta$ runs over $\{0,1,2,3\}$ such that $ \{\widetilde{\delta}_0^{(3)},\widetilde{\delta}_1^{(3)},\widetilde{\delta}_2^{(3)},\widetilde{\delta}_3^{(3)}\} = \left\{\delta^{(3)}, \delta\mathcal{G}^{(2)},\mathcal{G}^{(3)}, \Gamma^{(3)} \right\}$, respectively. In the main text, we expand the dot products shown above using \cite{Ortola_4PCF}:
\begin{align} \label{eq:exp_sph}
&(\mathbf{\widehat{a}}\cdot \mathbf{\widehat{b}})^{n} = \sum_{j=n,n-2,\cdots} \frac{n!(2j+1)}{2^{(n-j)/2}\left((1/2)\times(n-j)\right)!(j+n+1)!!}  \frac{4\pi}{2j+1}\sum_{m_j=-j}^{j} Y_{j,m_j}(\mathbf{\widehat{a}}) Y_{j,m_j}^{*}(\mathbf{\widehat{b}}) \nonumber \\
& \qquad \quad\;\;\equiv \sum_{j,m} d_{j}^{(n)} \;Y_{j,m_j}(\mathbf{\widehat{a}}) Y_{j,m_j}^{*}(\mathbf{\widehat{b}}),
\end{align}
where the result is obtained from writing the dot product as a sum over Legendre polynomials, and then the latter in terms of the spherical harmonics using the addition theorem. 

\section{Evaluation of the Angular integrals for the Covariance with Third-Order Densities}\label{Sec:Angular_Integrals}
\qquad In this section we show the evaluation of the main three angular integrals in this work. We start with the evaluation of one 3-argument isotropic basis function when all its arguments are the same: 
\begin{align}\label{eq:modified_Gaunt}
&\mathcal{G}_{L_1,L_2,L_3} = \int_{\hatk} \; \PP_{L_1,L_2,L_3}(\hatk,\hatk,\hatk)  \nonumber \\
&= (-1)^{L_1+L_2+L_3}\sum_{M_1,M_2,M_3}\tj{L_1}{L_2}{L_3}{M_1}{M_2}{M_3}\int_{\hatk} \;Y_{L_1,M_1}(\hatk)Y_{L_2,M_2}(\hatk)Y_{L_3,M_3}(\hatk) \nonumber \\
& = (-1)^{L_1+L_2+L_3}\left(\frac{(2L_1+1)(2L_2+1)(2L_3+1)}{4\pi}\right)^{1/2}\tj{L_1}{L_2}{L_3}{0}{0}{0}\sum_{M_1,M_2,M_3}\tj{L_1}{L_2}{L_3}{M_1}{M_2}{M_3}^2\nonumber \\
& = \left(\frac{(2L_1+1)(2L_2+1)(2L_3+1)}{4\pi}\right)^{1/2}\tj{L_1}{L_2}{L_3}{0}{0}{0}.
\end{align}
In the second equality, we have expanded the 3-argument isotropic basis function using its definition:
\begin{align}\label{eq:def_Iso_Basis}
&\mathcal{P}_{L,L',L"}(\hatk,\hatk,\hatk) = (-1)^{L+L'+L"}\sum_{M,M',M",m} \begin{pmatrix}
L & L' & L"\\
M & M' & M"
\end{pmatrix}\nonumber \\
& \qquad \qquad \qquad \qquad\times\int_{\hatk} Y_{L,M}(\hatk)  Y_{L',M'}(\hatk)  Y_{L",M"}(\hatk). 
\end{align}
In the third equality, we have used the known result of the Gaunt integral, given by \cite{DLMF} in Eq. (34.3.22):
\begin{align}\label{eq:Gaunt_Coeff}
   G_{M_1,M_2,M_3}^{L_1,L_2,L_3}\equiv \left(\frac{(2L_1+1)(2L_2+1)(2L_3+1)}{4\pi}\right)^{1/2}\tj{L_1}{L_2}{L_3}{0}{0}{0}\tj{L_1}{L_2}{L_3}{M_1}{M_2}{M_3},
\end{align}
and in the fourth equality we used Eq. 34.3.18 in \cite{DLMF}, and dropped the negative sign since the $3j$ symbol imposes an even sum of $L_1,\;L_2$ and $L_3$.  

We next compute the integral over one 2-argument isotropic basis function times two spherical harmonics when all the arguments are the same. We want the angular basis to be the same, so we expand the isotropic function into spherical harmonics and obtain:
\begin{align}\label{eq:Def_H_bar}
&\mathcal{\overline{H}}_{L,L',M,M'}^{\ell,\ell',m,m'} \equiv \int_{\hatk} Y_{L,m}(\hatk)Y_{L',m'}(\hatk)Y_{\ell,m}(\hatk)Y_{\ell',m'}(\hatk)\nonumber\\
&\qquad\quad\;\;= \sum_{\ell^",m^"} G_{M,M',-m^"}^{L,L',\ell^"}G_{m^",m,m'}^{\ell^",\ell,\ell'}.
\end{align}

Next, we evaluate the integral of a product of one 3-argument isotropic basis function and two spherical harmonics when all the arguments are the same:
\begin{align}\label{eq:Fancy_J_def}
&\mathcal{J}_{L,L',L^"}^{\ell,\ell',m,m^{'}} \equiv  \int_{\hatk} \mathcal{P}_{L,L',L^"}(\hatk,\hatk,\hatk) 
Y_{\ell,m}(\hatk) Y_{\ell',m'}(\hatk)  \nonumber \\
& \qquad \qquad\;\;\; = (-1)^{L+L'+L^"} \sum_{M,M',M^"} \begin{pmatrix}
L & L' & L^"\\
M & M' & M^"
\end{pmatrix}\nonumber \\
& \qquad \qquad \qquad  \times \int_{\hatk} Y_{L,M}(\hatk)  Y_{L',M'}(\hatk)  Y_{L^",M^"}(\hatk)  Y_{\ell,m}(\hatk) Y_{\ell',m'}(\hatk) \nonumber \\
& \qquad \qquad\;\;\;  = (-1)^{L+L'+L^"} \sum_{M,M',M^"}\;\sum_{\ell^",m^"}\sum_{\ell^{'''},m^{'''}} \begin{pmatrix}
L & L' & L^"\\
M & M' & M^"
\end{pmatrix} \nonumber \\
& \qquad \qquad \qquad  \times G_{L^", \ell, \ell^"}^{M^",m,-m^"} G_{\ell',\ell^", \ell'}^{m',m^",-m^{'''}} G_{L,L', \ell^{'''}}^{M,M',m^{'''}},
\end{align}
for which we have followed the same steps as in Eq. (\ref{eq:modified_Gaunt}) to obtain the result. To arrive at the third equality, we combined two spherical harmonics into a single harmonic twice, resulting in an integral with a product of three spherical harmonics. 

Finally, we evaluate the angular integral of a 3-argument isotropic basis function times four spherical harmonics when all the arguments are the same:
\begin{align}\label{eq:Fancy_L_def}
&\mathcal{L}_{L1,L2,L3}^{j_1,j_2,j_3,j_4,\{m_{j_i}\}}\equiv \int_{\hatk} \PP_{L1,L2,L3}(\hatk,\hatk,\hatk)\; Y_{j_1,m_1}(\hatk)Y_{j_2,m_2}(\hatk)Y_{j_3,m_3}(\hatk)Y_{j_4,m_4}(\hatk) \nonumber \\
& \qquad \qquad = (-1)^{L_1+L_2+L_3}\; \sum_{{\rm All}\;M } \sum_{\ell,\ell',\ell^",\ell^{(3)}}\tj{L_1}{L_2}{L_3}{M_1}{M_2}{M_3} G_{M_1,M_2,-m}^{L_1,L_2,\ell} \nonumber \\
& \qquad \qquad \quad \times G_{m,M_3,-m'}^{\ell,L_3,\ell'} G_{m_1,m_2,-m^"}^{j_1,j_2,\ell^"}G_{m_3,m_4,-m^{(3)}}^{j_3,j_4,\ell^{(3)}}G_{m^{(3)},m^",m'}^{\ell^{(3)},\ell^",\ell'},
\end{align}
where $\{m_{j_i}\} = {m_{j_1},m_{j_2},m_{j_3},m_{j_4}}$. In the second equality, we have expanded the isotropic basis functions into spherical harmonics, and then have combined two spherical harmonics recursively until the last three spherical harmonics remaining can be evaluated as a Gaunt integral.

\section{Evaluation of $\hats$ integrals}
In this section, we evaluate the $\hats$ integrals of the secondary-secondary, primary-secondary, and primary-primary configurations following \cite{Cahn_Iso, Hou_Cov} by averaging the isotropic basis over rotations, $R,\;R',$ and $S$. Averaging over rotations $R$ and $R'$ enables to express the angular dependence in terms of a single isotropic basis function, while averaging over rotation $S$ enables the evaluation of the $\hats$ dependence as previously demonstrated in \cite{Cahn_Iso, Hou_Cov}.
\subsection{$\hats$ Integral for the Secondary-Secondary Configuration}\label{sec:s_hat_SS_Evaluation}
We begin our analysis of these integrals with the secondary-secondary configuration:
\begin{align}
&\int dR\;dR'\;dS \;\PP_{L_{11}}(R\hatr_1,R\hatr_2)\PP_{L_{q20},L_{q2s},L'_{q21}}(R\hatr_0,-S\hats,-R'\hatr'_1) \PP_{L_{33},L'_{3s},L'_{33}}(R\hatr_3,-S\hats,-R'\hatr'_3)\nonumber \\
&\qquad\qquad\qquad \times \PP_{L_{q30},L_{q3s},L'_{q32}}(R\hatr_0,-S\hats,-R'\hatr'_2)\PP_{L_{q10},L_{q1s},L'_{q10}}(R\hatr_0,-S\hats,-R'\hatr'_0). 
\end{align}
Expanding the isotropic functions into spherical harmonics we find:
\begin{align}
&\int dR\;dR'\;dS\; (-1)^{L_{q10}+L_{q20}+L_{q30}+L_{33}} \sum_{{\rm All\;} M}\mathcal{C}^{L_{q20},L_{q2s},L'_{q21}}_{M_{q20},M_{q2s},M'_{q21}}\mathcal{C}^{L_{33},L'_{3s},L'_{33}}_{M_{33},M'_{3s},M'_{33}}\nonumber \\
&\qquad\times \mathcal{C}^{L_{q30},L_{q3s},L'_{q32}}_{M_{q30},M_{q3s},M'_{q32}}\mathcal{C}^{L_{q10},L_{q1s},L'_{q10}}_{M_{q10},M_{q1s},M'_{q10}} D_{L_{11}}Y_{L_{q10},M_{q10}}(R\hatr_0) Y_{L_{q20},M_{q20}}(R\hatr_0)\nonumber \\
&\qquad\times Y_{L_{q30},M_{q30}}(R\hatr_0)Y_{L_{11},M_{11}}(R\hatr_1)Y_{L_{11},M_{11}}^*(R\hatr_2)Y_{L_{33},M_{33}}(R\hatr_3)\nonumber \\
&\qquad\times Y_{L_{q1s},M_{q1s}}(S\hats)Y_{L_{q2s},M_{q2s}}(S\hats) Y_{L_{q3s},M_{q3s}}(S\hats)Y_{L'_{3s},M'_{3s}}(S\hats) \nonumber \\
&\qquad\times Y_{L'_{q10},M'_{q10}}(R'\hatr'_0) Y_{L'_{q21},M'_{q21}}(R'\hatr'_1)Y_{L'_{q10},M'_{q10}}(R'\hatr'_2) Y_{L'_{q21},M'_{q21}}(R'\hatr'_3), 
\end{align}
where we define $\mathcal{C}$ as \cite{Hou_Cov}:
\begin{align}
&\mathcal{C}^{\Lambda}_M = \sqrt{2\ell_{12}+1}\times \cdots\times\sqrt{2\ell_{12\cdots n-2}+1}\nonumber\\
&\qquad \quad \times \sum_{m_{12},\cdots}(-1)^\kappa\tj{\ell_1}{\ell_2}{\ell_{12}}{m_1}{m_2}{-m_{12}} \tj{\ell_{12}}{\ell_3}{\ell_{123}}{m_{12}}{m_3}{-m_{123}} \cdots\nonumber \\
&\qquad \quad \times \tj{\ell_{12\cdots n-2}}{\ell_{n-1}}{\ell_n}{m_{12\cdots n-2}}{m_{n-1}}{m_n},
\end{align}
and $\Lambda = \{\ell_1,\ell_2,(\ell_{12}),\ell_3,(\ell_{123}),\cdots,\ell_n\}$ with the 'intermediate' angular momenta in brackets, $M = \{m_1,m_2,(m_{12}),m_3,(m_{123}),\cdots,m_n\}$, and $\kappa=\ell_{12}-m_{12}+\ell_{123}-m_{123}+\cdots+\ell_{12\ldots n-2}-m_{12... n-2}$.

We analyze the integrals one by one, starting with the integral over rotations of $\hats$, for which we find:
\begin{align}
&\int dS\;Y_{L_{q1s},M_{q1s}}(S\hats)Y_{L_{q2s},M_{q2s}}(S\hats)s Y_{L_{q3s},M_{q3s}}(S\hats)Y_{L'_{3s},M'_{3s}}(S\hats) \nonumber \\
& = \int d\widehat{\mathbf{n}} \;Y_{L_{q1s},M_{q1s}}(\widehat{\mathbf{n}})Y_{L_{q2s},M_{q2s}}(\widehat{\mathbf{n}})s Y_{L_{q3s},M_{q3s}}(\widehat{\mathbf{n}})Y_{L'_{3s},M'_{3s}}(\widehat{\mathbf{n}}) \nonumber \\
& =\sum_{\ell,m} G_{M_{q2s},M'_{3s},-m}^{L_{q2s},L'_{3s},\ell}G_{m, M_{q3s},M_{q1s}}^{\ell,L_{q3s},L_{q1s}}, 
\end{align}
where we have defined $\widehat{\mathbf{n}}\equiv S\hats$ in the first equality, and written the resulting integral in the second equality in terms of Gaunt integrals. 

Next, we evaluate the integral over the rotational average of the $\hatr'$ vectors, for which the solution is explicitly given in Eq. (2.31) of \cite{Cahn_Iso}. We state the result here:  
\begin{align}\label{eq:R_prime_integral}
&\int dR'\;Y_{L'_{q10},M'_{q10}}(R'\hatr'_0) Y_{L'_{q21},M'_{q21}}(R'\hatr'_1) Y_{L'_{q32},M'_{q32}}(R'\hatr'_2) Y_{L'_{33},M'_{33}}(R'\hatr'_3)\nonumber \\
&= \sum_{\Lambda'} \mathcal{C}^{\Lambda'}_{M'}\PP_{\Lambda'}(\hatR'), 
\end{align}
where we have defined $\Lambda'=\left\{L'_{q10},L'_{q21},L'_{q32},L'_{33}\right\}$ and likewise for $M'$. 

Finally, we evaluate the integral over the rotational average of the $\hatr$:
\begin{align}\label{eq:R_integral}
&\int dR\;Y_{L_{q10},M_{q10}}(R\hatr_0) Y_{L_{q20},M_{q20}}(R\hatr_0) Y_{L_{q30},M_{q30}}(R\hatr_0)\nonumber \\
& \times Y_{L_{11},M_{11}}(R\hatr_1)Y_{L_{11},M_{11}}^*(R\hatr_2) Y_{L_{33},M_{33}}(R\hatr_3)\nonumber \\
&= \sum_{\ell',\ell^"}\sum_{m',m^"} G^{L_{q20},L_{q30},\ell'}_{M_{q20},M_{q30},-m'} G^{\ell',L_{q10},\ell^"}_{m',M_{q10},-m^"}\nonumber \\
&  \times\int dR\; Y_{\ell^",m^"}(R\hatr_0)Y_{L_{11},M_{11}}(R\hatr_1)Y_{L_{11},M_{11}}^*(R\hatr_2)Y_{L_{33},M_{33}}(R\hatr_3)\nonumber \\
&=(-1)^{M_{11}}\sum_{\Lambda}\mathcal{C}_{M}^{\Lambda}\PP_{\Lambda}(\hatR) \sum_{\ell',m'}G^{L_{q20},L_{q30},\ell'}_{M_{q20},M_{q30},-m'} G^{\ell',L_{q10},\ell^"}_{m',M_{q10},-m^"},
\end{align}
where we have combined the three spherical harmonics of the $\hatr_0$ into a single one in the second equality, and in the last equality we have used Eq. (2.31) of \cite{Cahn_Iso} once again. We have also defined $\Lambda \equiv\left\{\ell^",L_{11},L_{11},L_{33}\right\}$.

The result of the $\hats$ integral for the secondary-secondary configuration can be written as:
\begin{align}\label{eq:s_hat_SS}
&\int dR\;dR'\;dS \;\PP_{L_{11}}(R\hatr_1,R\hatr_2)\PP_{L_{q20},L_{q2s},L'_{q21}}(R\hatr_0,-S\hats,-R'\hatr'_1) \PP_{L_{33},L'_{3s},L'_{33}}(R\hatr_3,-S\hats,-R'\hatr'_3)\nonumber \\
&\qquad\qquad\qquad \times \PP_{L_{q30},L_{q3s},L'_{q32}}(R\hatr_0,-S\hats,-R'\hatr'_2)\PP_{L_{q10},L_{q1s},L'_{q10}}(R\hatr_0,-S\hats,-R'\hatr'_0)\nonumber \\
& = \sum_{\Lambda,\Lambda_s,\Lambda'}\overline{\mathcal{Q}}_{(\rm S-S)}^{\Lambda,\Lambda_s,\Lambda'} \;\PP_{\Lambda} (\hatR)\PP_{\Lambda'}(\hatR'), 
\end{align}
where $\Lambda_s =\left\{L_{q1s},L_{q2s},L_{q3s},L'_{3s}\right\}$ and $\overline{\mathcal{Q}}$ has been defined as:
\begin{align}\label{eq:Q_bar_def_SS}
&\overline{\mathcal{Q}}_{(\rm S-S)}^{\Lambda,\Lambda_s,\Lambda'} \equiv \sum_{\rm All\; M}(-1)^{M_{11}+L_{q20}+L_{33}+L_{q30}+L_{q10}}D_{L_{11}}\mathcal{C}^{L_{q20},L_{q2s},L'_{q21}}_{M_{q20},M_{q2s},M'_{q21}}\mathcal{C}^{L_{33},L'_{3s},L'_{33}}_{M_{33},M'_{3s},M'_{33}} \nonumber \\
& \qquad \qquad \times\mathcal{C}^{L_{q30},L_{q3s},L'_{q32}}_{M_{q30},M_{q3s},M'_{q32}}\mathcal{C}^{L_{q10},L_{q1s},L'_{q10}}_{M_{q10},M_{q1s},M'_{q10}} \mathcal{C}^{\Lambda'}_{M'}\mathcal{C}^{\Lambda}_{M}\nonumber \\
& \qquad \qquad \times \sum_{\ell,\ell'} \sum_{m,m'} G_{M_{q2s},M'_{3s},-m}^{L_{q2s},L'_{3s},\ell}G_{m, M_{q3s},M_{q1s}}^{\ell,L_{q3s},L_{q1s}}G^{L_{q20},L_{q30},\ell'}_{M_{q20},M_{q30},-m'} G^{\ell',L_{q10},L_{33}}_{m',M_{q10},-M_{33}}.
\end{align}

\subsection{$\hats$ Integral for the Primary-Secondary Configuration}\label{sec:s_hat_PS_Evaluation}
We continue the analysis of the $\hats$ integrals by evaluating the Primary-Secondary configurations:
\begin{align}
&\int dR\;dR'\;dS \;Y_{L_{q10},M_{q10}}^*(\hatr_0)Y_{L_{q11},M_{q11}}(\hatr_1)\PP_{L_{q20},L_{q2s},L'_{q20}}(R\hatr_0,-S\hats,-R'\hatr'_0) \nonumber \\
&\times \PP_{L_{q30},L_{q3s},L'_{q31}}(R\hatr_0,-S\hats,-R'\hatr'_1)\PP_{L_{22},L'_{2s},L'_{22}}(R\hatr_2,-S\hats,-R'\hatr'_2)\nonumber \\
&\times\PP_{L_{33},L'_{3s},L'_{33}}(R\hatr_3,-S\hats,-R'\hatr'_3). 
\end{align}
The result of the $dS$ integral will follow the same result as in the previous section with the only difference that $\Lambda_s = \left\{L_{q2s},L_{q3s},L'_{2s},L'_{3s}\right\}$; likewise the $dR'$ integral evaluates to the same result as in the previous section with $\Lambda' = \left\{L'_{q20},L'_{q31},L'_{22},L'_{33}\right\}$. 

Therefore, we only need to evaluate the $dR$ integral:
\begin{align}
&\int dR\;Y_{L_{q20},M_{q20}}(R\hatr_0) Y_{L_{q30},M_{q30}}(R\hatr_0) Y_{L_{q10},M_{q10}}^*(R\hatr_0)\nonumber \\
&\qquad \times Y_{L_{q11},M_{q11}}(R\hatr_1)Y_{L_{22},M_{22}}(R\hatr_2)Y_{L_{33},M_{33}}(R\hatr_3)\nonumber \\
&= \sum_{\ell',m'} G^{L_{q20},L_{q30},\ell'}_{M_{q20},M_{q30},-m'}G^{L_{q10},\ell',\ell^"}_{-M_{q10},m',-m^"} \nonumber \\
&\;\;\;\times\int dR\; Y_{\ell^",m^"}(R\hatr_0)Y_{L_{q10},M_{q10}}(R\hatr_1)Y_{L_{22},M_{22}}(R\hatr_2)Y_{L_{33},M_{33}}(R\hatr_3)\nonumber \\
&=\sum_{\Lambda}\mathcal{C}^{\Lambda}_{M}\PP_{\Lambda}(\hatR) \sum_{\ell',\ell^"}\sum_{m',m^"}G^{L_{q20},L_{q30},\ell'}_{M_{q20},M_{q30},-m'}G^{L_{q10},\ell',\ell^"}_{-M_{q10},m',-m^"},
\end{align}
where we have defined $\Lambda \equiv \left\{\ell^",L_{q11},L_{22},L_{33}\right\}$. Thus, the result of evaluating the $\hats$ integral by averaging over rotations results in:
\begin{align}\label{eq:s_hat_PP_result}
&\int dR\;dR'\;dS \;\PP_{L_{q10}}(\hatr_0,\hatr_1)\PP_{L_{q20},L_{q2s},L'_{q20}}(R\hatr_0,-S\hats,-R'\hatr'_0) \PP_{L_{q30},L_{q3s},L'_{q31}}(R\hatr_0,-S\hats,-R'\hatr'_1)\nonumber \\
&\qquad\qquad\qquad \times \PP_{L_{22},L'_{2s},L'_{22}}(R\hatr_2,-S\hats,-R'\hatr'_2)\PP_{L_{33},L'_{3s},L'_{33}}(R\hatr_3,-S\hats,-R'\hatr'_3)\nonumber \\
&= \sum_{\Lambda,\Lambda_s,\Lambda'} \overline{\mathcal{Q}}_{(\rm P-S)}^{\Lambda,\Lambda_s,\Lambda'}\PP_{\Lambda}(\hatR)\PP_{\Lambda'}(\hatR'),
\end{align}
where we have defined:
\begin{align}\label{eq:Q_bar_def_PS}
&\overline{\mathcal{Q}}_{(\rm P-S)}^{\Lambda,\Lambda_s,\Lambda'} \equiv \sum_{\rm All\; M}(-1)^{L_{q20}+L
_{q30}+L_{22}+L_{33}}\;D_{L_{q10}}\mathcal{C}^{\Lambda}_{M}\;\mathcal{C}^{\Lambda'}_{M'}\;\mathcal{C}^{L_{q20},L_{q2s},L'_{q20}}_{M_{q20},M_{q2s},M'_{q20}}\mathcal{C}^{L_{q30},L_{q3s},L'_{q31}}_{M_{q30},M_{q3s},M'_{q31}}  \nonumber \\
& \quad \qquad \quad \times \mathcal{C}^{L_{22},L'_{2s},L'_{22}}_{M_{22},M'_{2s},M'_{22}}\mathcal{C}^{L_{33},L'_{3s},L'_{33}}_{M_{33},M'_{3s},M'_{33}} \sum_{\ell',\ell^"}\sum_{m',m^"}G^{L_{q20},L_{q30},\ell'}_{M_{q20},M_{q30},-m'}G^{L_{q10},\ell',\ell^"}_{-M_{q10},m',-m^"}.
\end{align}
\subsection{$\hats$ Integral for the Primary-Primary Configuration}\label{sec:s_hat_PP_Evaluation}
We continue by evaluating the $\hats$ integral for the primary-primary configuration:
\begin{align}
&\int dR\;dR'\;dS \;\PP_{L_{00},L_{0s},L'_{00}}(R\hatr_0,-S\hats,-R'\hatr'_0) \PP_{L_{11},L_{1s},L'_{11}}(R\hatr_1,-S\hats,-R'\hatr'_1)\nonumber \\
&\qquad\qquad\qquad \times \PP_{L_{22},L'_{2s},L'_{22}}(R\hatr_2,-S\hats,-R'\hatr'_2)\PP_{L_{33},L'_{3s},L'_{33}}(R\hatr_3,-S\hats,-R'\hatr'_3). 
\end{align}
This integral has already been solved in \cite{Hou_Cov, Cahn_Iso}, so we simply state the result:
\begin{align}
&\int dR\;dR'\;dS \;\PP_{L_{00},L_{0s},L'_{00}}(R\hatr_0,-S\hats,-R'\hatr'_0) \PP_{L_{11},L_{1s},L'_{11}}(R\hatr_1,-S\hats,-R'\hatr'_1)\nonumber \\
&\qquad\qquad\qquad \times \PP_{L_{22},L'_{2s},L'_{22}}(R\hatr_2,-S\hats,-R'\hatr'_2)\PP_{L_{33},L'_{3s},L'_{33}}(R\hatr_3,-S\hats,-R'\hatr'_3)\nonumber \\
&\qquad\qquad\qquad = (4\pi)^{-2} \sum_{\Lambda,\Lambda_s,\Lambda'}\overline{\mathcal{Q}}_{\rm (P-P)}^{\Lambda,\Lambda_s,\Lambda'}\mathcal{S}^{P}_{\Lambda_s}\mathcal{C}^{\Lambda_s}_{\mathbf{0}}\; \PP_{\Lambda}(\hatR)\PP_{\Lambda'}(\hatR'),
\end{align}
where $\Lambda = \left\{L_{00},L_{11},L_{22},L_{33} \right\}$, $\Lambda_s = \left\{L_{0s},L_{1s},L_{2s},L_{3s} \right\}$ and $\Lambda' = \left\{L'_{00},L'_{11},L'_{22},L'_{33} \right\}$. The constants have been defined as:
\begin{align}
\overline{\mathcal{Q}}_{\rm (P-P)}^{\Lambda,\Lambda_s,\Lambda'}\equiv\prod_{i=1}^n\sum_{M_{ii} M_{is} M'_{ii}}\mathcal{C}^{L_{ii}L_{is} L'_{ii}}_{M_{ii} M_{is} M'_{ii}}
\mathcal{C}^{\Lambda}_{\rm M} \mathcal{C}^{\Lambda_s}_{\rm M_s} \mathcal{C}^{\Lambda'}_{\rm M'},
\end{align}
and
\begin{align}
    \mathcal{S}^{P}_{\Lambda} \equiv \prod_{j=1}^{N}s_{\ell_j}^{(\rm I)}. 
\end{align}

\section{Forming the Connected 4PCF Covariance}\label{sec:Forming_conn_4PCF}
We show how to form the connected 4PCF covariance terms that enter at 1-loop order with the third-order densities. Following \cite{Philcox_4PCF_measurement}, every integral over correlation functions scales as $r_{\text{c}}^3/V$, where $r_{\text{c}} \approx 100$ Mpc/$h$is the correlation length and $V$ is the volume of the survey. Because the survey size tends to be much larger than the correlation length, for every integral with correlation functions, we get a suppression of the order of $\left(r_{\text{c}}^3/V\right)^{\alpha}$, where $\alpha = \left\{1,2,3,\cdots\right\}$ indicates the number of integrals over correlation functions we have. 

Therefore, we compute the connected 4PCF covariance with only the terms that are of the same order in $r_{\text{c}}^3/V$. We first define the components that give rise to the connected 4PCF, starting with the full 4PCF:
\begin{align}
&\hat\zeta^{(\rm full)}(\vecr_1,\vecr_2,\vecr_3) = \lim_{\vecr_0 \to 0}\frac{1}{V}\int d^3\vecx\,\delta(\vecx+\vecr_0)\delta(\vecx+\vecr_1)\delta(\vecx+\vecr_2)\delta(\vecx+\vecr_3).
\end{align}
Here, $\vecr_0$ is introduced for symmetry purposes, as in $\S$\ref{sec:Cov_Compu}, and is taken to zero at the end. For brevity, we omit this limit in what follows. Next, we continue with the disconnected piece:
\begin{align}
&\hat\zeta^{(\rm disc.)}(\vecr_1,\vecr_2,\vecr_3) = \frac{1}{V}\int d^3\vecx_1\,\delta(\vecx_1+\vecr_0)\delta(\vecx_1+\vecr_1)\;\frac{1}{V}\int d^3\vecx_2 \;\delta(\vecx_2+\vecr_2)\delta(\vecx_2+\vecr_3). 
\end{align}
Given the above definitions, to compute the connected 4PCF covariance we must then obtain:
\begin{align}\label{eq:Connected_Cov_definition}
&\mathrm{Cov}_{4,\text{1L}}^{[3],( c, c)} = \mathrm{Cov}_{4,\text{1L}}^{[3],(\rm full, full)} - \mathrm{Cov}_{4,\text{1L}}^{[3],(\rm full, disc)} - \mathrm{Cov}_{4,\text{1L}}^{[3],(\rm disc, full)} + \mathrm{Cov}_{4,\text{1L}}^{[3],(\rm disc, disc)}. 
\end{align}
We show how to calculate $\mathrm{Cov}^{(\rm full, full)}$ with the third-order density contrast, but the other terms follow the same procedure:
\begin{align}
&\mathrm{Cov}^{(\rm full, full)} = \left\langle \frac{1}{V}\int d^3\vecx\,\delta^{(3)}(\vecx+\vecr_0)\dlin(\vecx+\vecr_1)\dlin(\vecx+\vecr_2)\dlin(\vecx+\vecr_3)\right.\nonumber \\
& \qquad \qquad \qquad \left. \times \frac{1}{V}\int d^3\vecx'\,\dlin(\vecx'+\vecr'_0)\dlin(\vecx'+\vecr'_1)\dlin(\vecx'+\vecr'_2)\dlin(\vecx'+\vecr'_3)\right\rangle.
\end{align}
We expand the third-order density in terms of three linear densities in real space\footnote{We set the third-order kernel to unity just for the demonstration of how to construct the connected 4PCF covariance. In the main text, we evaluate the covariance with the third-order kernel incorporated.} and apply Wick's theorem to obtain:
\begin{align}\label{eq:Cov_FullxFull_term}
&\mathrm{Cov}_{4,\text{1L}}^{[3],(\rm full, full)} \supset \frac{1}{V^2}\int d^3\vecq_1\;d^3\vecq_2\;d^3\vecq_3\int d^3\vecx\int d^3\vecx'\left[\left\{ \left\langle \dlin(\vecx+\vecr_0+\vecq_1) \dlin (\vecx+\vecr_1) \right\rangle \right. \right. \nonumber \\
& \qquad \qquad \qquad\left.\left. \times \left\langle \dlin(\vecx+\vecr_0+\vecq_2) \dlin (\vecx'+\vecr'_0) \right\rangle \left\langle \dlin(\vecx+\vecr_0+\vecq_3) \dlin (\vecx'+\vecr'_1) \right\rangle \right. \right. \nonumber \\
& \qquad \qquad \qquad\left. \left. \times\left\langle \dlin(\vecx+\vecr_2) \dlin (\vecx'+\vecr'_2) \right\rangle \left\langle \dlin(\vecx+\vecr_3) \dlin (\vecx'+\vecr'_3) \right\rangle + \text{215 perms.}\footnotemark[5] \right\} \right. \nonumber \\
& \qquad \qquad \qquad  + \left\{ \left. \left\langle \dlin(\vecx+\vecr_0+\vecq_1) \dlin (\vecx+\vecr_1) \right\rangle \left\langle \dlin(\vecx+\vecr_0+\vecq_2) \dlin (\vecx+\vecr_2) \right\rangle\right. \right. \nonumber \\ 
& \qquad \qquad \qquad\left. \left. \times \left\langle \dlin(\vecx+\vecr_0+\vecq_3) \dlin (\vecx'+\vecr'_0) \right\rangle \left\langle \dlin(\vecx'+\vecr'_1) \dlin (\vecx'+\vecr'_2) \right\rangle\right.\right. \nonumber \\
& \qquad \qquad \qquad\left. \left. \times \left\langle \dlin(\vecx+\vecr_3) \dlin (\vecx'+\vecr'_3) \right\rangle + \text{359 perms.}\footnotemark[6] \right\} \right] \nonumber \\
& \qquad \qquad \;\; = \frac{1}{V}\int d^3\vecq_1\;d^3\vecq_2\;d^3\vecq_3\int d^3\vecs \left[ \left\{\xi_W( \vecr_0-\vecr_1+\vecq_1)\xi_W( \vecr_0-\vecs-\vecr'_0+\vecq_2)\right.\right. \nonumber \\
& \qquad \qquad \qquad\left.\left. \times\xi( \vecr_0-\vecs-\vecr'_1+\vecq_3)\xi( \vecr_2-\vecs-\vecr'_2)\xi( \vecr_3-\vecs-\vecr'_3)+ \text{215 perms.}\right\} \right. \nonumber \\
& \qquad \qquad \qquad  + \left. \left\{ \xi_W(\vecr_0-\vecr_1+\vecq_1)\xi_W(\vecr_0-\vecr_2+\vecq_2)\xi_W(\vecr'_0-\vecr'_0+\vecq_3)\right.\right. \nonumber \\
& \qquad \qquad \qquad\left. \left.\times\xi( \vecr_2-\vecs-\vecr'_2)\xi( \vecr_3-\vecs-\vecr'_3)+ \text{359 perms.}\right\}\right],
\end{align}
\footnotetext[5]{The number of permutation has been obtained by adding the permutations counted in the main text for the fully-coupled terms.}\footnotetext[6]{Finding the number of permutations is done in three steps. i) Pair one linear density with another one in the tetrahedron with no third-order density; one has 6 possibilities. ii) Pair the remaining two linear densities with two of the six densities in the third-order tetrahedron; one has 30 possibilities. iii) Pair the remaining four densities in the third-order tetrahedron; one has two possibilities. Total number of permutations = 6 $\times$ 30$\times$ 2 = 360 perms.}where we have expressed the correlation function derived from the third-order densities as $\xi_W$ to indicate that this must actually be evaluated with their respective third-order kernels as shown in the main text. We have also defined $\vecs=\vecx'-\vecx$. The correlation functions in the line before last arise from contracting density contrasts within the same tetrahedrons; we term these structures partially-connected correlations. All of the above permutations are of the order $r_\text{c}^3/V$. 

Using Eq. (\ref{eq:Connected_Cov_definition}), all the terms with the same order, $r_\text{c}^3/V$, will cancel each other. This means we must check the full $\times$ disconnected, disconnected $\times$ full, and disconnected $\times$ disconnected pieces carefully to determine which terms enter the connected 4PCF covariance at the 1-loop level. We find that all the partially-connected terms cancel and do not enter the covariance.

Below, we show an example of one of the partially-connected permutations that is being suppressed in the full $\times$ disconnected piece and disconnected $\times$ disconnected. However, this term is not suppressed in the disconnected $\times$ full, therefore canceling itself with the full $\times$ full:
\begin{align}
&\mathrm{Cov}_{4,\text{1L}}^{[3],(\rm full, disc.)} \supset \frac{1}{V^3}\int d^3\vecq_1\;d^3\vecq_2\;d^3\vecq_3\int d^3\vecx'\int d^3\vecx_1\int d^3\vecx_2\left[\left\langle \dlin(\vecx_1+\vecr_0+\vecq_1) \dlin (\vecx_1+\vecr_1) \right\rangle \right. \nonumber \\ 
& \qquad \qquad \qquad\left.  \times \left\langle \dlin(\vecx_1+\vecr_0+\vecq_2) \dlin (\vecx_2+\vecr_2) \right\rangle \left\langle \dlin(\vecx_1+\vecr_0+\vecq_3) \dlin (\vecx'+\vecr'_0) \right\rangle \right. \nonumber \\
& \qquad \qquad \qquad \left. \times \left\langle \dlin(\vecx'+\vecr'_1) \dlin (\vecx'+\vecr'_2) \right\rangle\left\langle \dlin(\vecx_2+\vecr_3) \dlin (\vecx'+\vecr'_3) \right\rangle \right] \nonumber \\
& \qquad \qquad \;\; = \frac{1}{V^3}\int d^3\vecq_1\;d^3\vecq_2\;d^3\vecq_3\int d^3\vecs\int d^3\vecs_1\int d^3\vecs_2\left[ \xi_W(\vecr_0-\vecr_1+\vecq_1)\xi_W(\vecr_0-\vecs-\vecr_2+\vecq_2)\right. \nonumber \\
& \qquad \qquad \qquad\left. \times\xi_W(\vecr_0-\vecs_1-\vecr'_0+\vecq_3)\xi( \vecr'_2-\vecr'_1)\xi(\vecr_3-\vecs_2-\vecr'_3)\right],
\end{align}
where we defined $\vecs = \vecx_2 - \vecx_1$, $\vecs_1=\vecx'-\vecx_1$ and $\vecs_2=\vecx'-\vecx_2$. This term is being suppressed by $r_\text{c}^9/V^3$ and we next show the disconnected $\times$ full piece, at $r_\text{c}^3/V$:
\begin{align}
&\mathrm{Cov}_{4,\text{1L}}^{[3],(\rm disc., full)} \supset \frac{1}{V^3}\int d^3\vecq_1\;d^3\vecq_2\;d^3\vecq_3\int d^3\vecx \int d^3\vecx'_1\int d^3\vecx'_2\left[\left\langle \dlin(\vecx+\vecr_0+\vecq_1) \dlin (\vecx+\vecr_1) \right\rangle \right. \nonumber \\ 
& \qquad \qquad \qquad\left.  \times \left\langle \dlin(\vecx+\vecr_0+\vecq_2) \dlin (\vecx+\vecr_2) \right\rangle \left\langle \dlin(\vecx+\vecr_0+\vecq_3) \dlin (\vecx_1'+\vecr'_0) \right\rangle \right. \nonumber \\
& \qquad \qquad \qquad \left. \times \left\langle \dlin(\vecx_2'+\vecr'_1) \dlin (\vecx'_2+\vecr'_2) \right\rangle\left\langle \dlin(\vecx+\vecr_3) \dlin (\vecx'_1+\vecr'_3) \right\rangle \right] \nonumber \\
& \qquad \qquad \;\; = \frac{1}{V}\int d^3\vecq_1\;d^3\vecq_2\;d^3\vecq_3\int d^3\vecs\left[ \xi_W(\vecr_0-\vecr_1+\vecq_1)\xi_W(\vecr_0-\vecr_2+\vecq_2)\right. \nonumber \\
& \qquad \qquad \qquad\left. \times\xi_W(\vecr_0-\vecs-\vecr'_0+\vecq_3)\xi( \vecr'_2-\vecr'_1)\xi(\vecr_3-\vecs-\vecr'_3)\right],
\end{align}
where we have defined $\vecs = \vecx'_1-\vecx$. Finally, the disconnected $\times$ disconnected piece is obtained as:
\begin{align}
&\mathrm{Cov}_{4,\text{1L}}^{[3],(\rm disc., disc.)} \supset \frac{1}{V^4}\int d^3\vecq_1\;d^3\vecq_2\;d^3\vecq_3\int d^3\vecx_1 \int d^3\vecx_2 \nonumber \\
& \qquad \qquad \qquad \times \int d^3\vecx'_1\int d^3\vecx'_2\left[\left\langle \dlin(\vecx_1+\vecr_0+\vecq_1) \dlin (\vecx_1+\vecr_1) \right\rangle \right. \nonumber \\ 
& \qquad \qquad \qquad\left.  \times \left\langle \dlin(\vecx_1+\vecr_0+\vecq_2) \dlin (\vecx_2+\vecr_2) \right\rangle \left\langle \dlin(\vecx_1+\vecr_0+\vecq_3) \dlin (\vecx_1'+\vecr'_0) \right\rangle \right. \nonumber \\
& \qquad \qquad \qquad \left. \times \left\langle \dlin(\vecx_2'+\vecr'_1) \dlin (\vecx'_2+\vecr'_2) \right\rangle\left\langle \dlin(\vecx_2+\vecr_3) \dlin (\vecx'_1+\vecr'_3) \right\rangle \right] \nonumber \\
& \qquad \qquad \;\; = \frac{1}{V^3}\int d^3\vecq_1\;d^3\vecq_2\;d^3\vecq_3\int d^3\vecs \int d^3\vecs_1\int d^3\vecs_{12}\left[ \xi_W(\vecr_0-\vecr_1+\vecq_1)\xi_W(\vecr_0-\vecs_{12}-\vecr_2+\vecq_2)\right. \nonumber \\
& \qquad \qquad \qquad\left. \times\xi_W(\vecr_0-\vecs_1-\vecr'_0+\vecq_3)\xi( \vecr'_2-\vecr'_1)\xi(\vecr_3-\vecs-\vecr'_3)\right],
\end{align}
where we defined $\vecs = \vecx'_1 - \vecx_2$, $\vecs_1=\vecx_1'-\vecx_1$ and $\vecs_2=\vecx_1-\vecx_2$. This term is being suppressed at $r_\text{c}^9/V^3$. As anticipated above, when the above 4 solutions are substituted into Eq. (\ref{eq:Connected_Cov_definition}), all the partially-connected terms cancel out.

Consequently, after accounting for all the terms that cancel out and the terms that get suppressed, we find the connected 4PCF covariance at the 1-loop order for the third-order density contrast is:
\begin{align}
 &\mathrm{Cov}_{4,\text{1L}}^{[3],( c, c)} = \frac{1}{V}\int d^3\vecq_1\;d^3\vecq_2\;d^3\vecq_3\int d^3\vecs \left[ \xi_W( \vecr_0-\vecr_1+\vecq_1)\xi_W( \vecr_0-\vecs-\vecr'_0+\vecq_2)\right. \nonumber \\
& \qquad \qquad \qquad\left. \times\xi( \vecr_0-\vecs-\vecr'_1+\vecq_3)\xi( \vecr_2-\vecs-\vecr'_2)\xi( \vecr_3-\vecs-\vecr'_3)+ \text{215 perms.} \right]. 
\end{align}

\section{Next-to-Leading-Order Covariance Corrections Scales}
\label{sec:Correction_Scale}
The NLO contribution to the 4PCF covariance matrix is suppressed relative to the Gaussian contribution by two additional powers of the linear density contrast,
\begin{equation}
\label{eq:NLO_scaling_delta}
\mathrm{Cov}_{\mathrm{NLO}} \sim \mathcal{O}\!\left(\delta_{\rm lin}^2\right)\times\mathrm{Cov}_{\mathrm{G}} .
\end{equation}
This scaling follows from standard perturbation theory, where the Gaussian covariance is sourced by disconnected products of two-point functions, while the NLO covariance receives contributions from connected diagrams involving higher-order mode couplings.

To make this statement quantitative, we estimate the variance of the linear density field smoothed on a physical scale \(R\),
\begin{equation}
\sigma^2(R) \equiv \left\langle \delta_R^2 \right\rangle ,
\end{equation}
and use it as a proxy for the expected relative size of NLO corrections. In this approach, values of \(\sigma^2(R) = \{0.1, 0.2, \ldots\}\) correspond to scales at which the NLO covariance is expected to contribute at the \(\{10\%, 20\%, \ldots\}\) level relative to the Gaussian term.

We define the smoothed density contrast \(\delta_R\) via a real-space spherical top-hat filter of radius \(R\),
\begin{equation}
\delta_R(\mathbf{x}) \equiv \frac{1}{V_R} \int_{|\mathbf{y}|\le R} d^3y \,
\delta_{\rm lin}(\mathbf{x}+\mathbf{y}),
\qquad
V_R = \frac{4\pi R^3}{3}.
\end{equation}
For a statistically homogeneous and isotropic density field, the corresponding variance may be written in Fourier space as
\begin{equation}
\label{eq:sigma2_general}
\sigma^2(R)
= \int \frac{d^3k}{(2\pi)^3}\,
P_{\rm lin}(k)\, \left| W_{\rm TH}(kR) \right|^2 ,
\end{equation}
where \(P_{\rm lin}(k)\) is the linear matter power spectrum and
\begin{equation}
W_{\rm TH}(kR) = 3\,\frac{j_1(kR)}{kR}
\end{equation}
is the Fourier transform of the real-space top-hat window, with \(j_1\) the spherical Bessel function of order one.

Using isotropy to reduce Eq.~\eqref{eq:sigma2_general} to a one-dimensional integral, we obtain
\begin{equation}
\label{eq:sigma2_final}
\sigma^2(R)
= \frac{9}{2\pi^2}
\int_0^\infty dk\, k^2\,
P_{\rm lin}(k)\,
\left(\frac{j_1(kR)}{kR}\right)^2 .
\end{equation}

\begin{figure}[h!]
    \centering
    \includegraphics[width=0.8\linewidth]{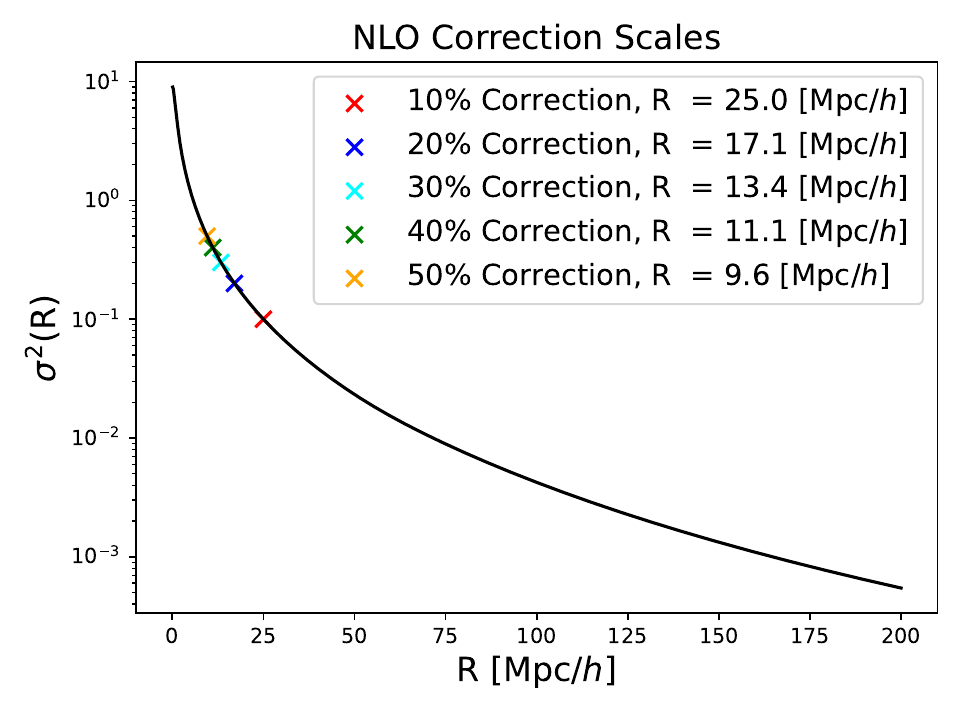}
    \caption{Estimate of the physical scales at which NLO corrections to the 4PCF covariance are expected to become significant. Shown is the variance of the linear density field, $\sigma^2(R)=\langle \delta_R^2\rangle$; see Eq.~(\ref{eq:sigma2_final}). The legend indicates the $10\%$--$50\%$ correction levels, together with their corresponding scales and the markers used in the figure.}
    \label{fig:NLO_Correction_Scales}
\end{figure}

We evaluate Eq.~\eqref{eq:sigma2_final} numerically using the same linear power spectrum employed throughout this work. By solving for the values of \(R\) at which \(\sigma^2(R)\) reaches fixed thresholds, we identify the characteristic scales at which the NLO covariance is expected to induce fractional corrections of \(10\%\), \(20\%\), \(30\%\), \(40\%\), and \(50\%\) relative to the Gaussian covariance. The resulting scales are shown in Fig.~\ref{fig:NLO_Correction_Scales}.

We emphasize that this procedure provides an order-of-magnitude estimate rather than a sharp transition scale. Nonetheless, it offers a physically motivated and transparent criterion for assessing the regime of validity of the Gaussian covariance approximation for the 4PCF, and for identifying the scales on which NLO corrections are expected to become non-negligible.

\section{Computational Cost: Brute Force vs. Analytical Reduction}\label{sec:Computational_Cost}
In this section, we estimate the computational cost of evaluating the next-to-leading order third-order covariance of the 4PCF using a direct brute-force approach, and contrast it with the cost of our analytical reduction. The goal is to show that a naive numerical evaluation of the covariance expression is computationally prohibitive, while the reduced formulation is tractable in practice.

We begin by considering a brute-force (BF) example (without loss of generality) using the secondary-secondary configuration shown in Fig. \ref{fig:dc_configurations}, schematically written in configuration space as:
\begin{align}
{\rm BF} \propto\bigg< \int \frac{d^3\vecs}{V}\; \delta^{(3)}(\vecx + \vecr_{0})  \prod_{p=1}^{3} \prod_{t=0}^{3} \delta_{\rm lin.}(\vecx+\vecr_p)\delta_{\rm lin.}(\vecx+\vecr'_t+ \vecs) \bigg>.
\end{align}
Expressing the third-order density fields in Fourier space and expanding all linear density fields, the above expression becomes:
\begin{align}
&{\rm BF} \propto \bigg< \int_{\vecs}\;\int_{\veck_0}e^{-i\veck_{0}\cdot(\vecx+\vecr_0)}\int d^3\vecq_1\;d^3\vecq_2\;d^3\vecq_3 \;\dD(\veck_0-\vecq_1-\vecq_2-\vecq_3) \nonumber \\
& \qquad \quad \times F^{[3]}(\vecq_1,\vecq_2,\vecq_3) \;\dlin(\vecq_1)\dlin(\vecq_2)\dlin(\vecq_3) \nonumber \\
& \qquad\quad \times \prod_{p} \prod_{t}\int_{\veck_p}e^{-i\veck_{p}\cdot(\vecx+\vecr_p)} \dlin(\veck_p) \int_{\veck'_{t}}e^{-i\veck'_{t}\cdot(\vecx+\vecr'_p+\vecs)} \dlin(\veck'_t) \bigg>.
\end{align}
After performing the ensemble average using Wick’s theorem and carrying out the trivial integrations, the expression reduces to:
\begin{align}
&{\rm BF} \propto \int_{\vecs} \int d^3\vecq_1\;d^3\vecq_2\;d^3\vecq_3 \;e^{-i\vecq_1\cdot(\vecr_0-\vecr'_0-\vecs)}\;e^{-i\vecq_2\cdot(\vecr_0-\vecr'_1-\vecs)} \;e^{-i\vecq_3\cdot(\vecr_0-\vecr'_2-\vecs)}\nonumber \\
& \qquad \times F^{[3]}(\vecq_1,\vecq_2,\vecq_3)\;\Pk(q_1)\Pk(q_2)\Pk(q_3)  \int_{\veck_1} \;e^{-i\veck_1\cdot(\vecr_1-\vecr_2)} \Pk(k_1)\nonumber \\
& \qquad \times \int_{\veck_3} \;e^{-i\veck_3\cdot(\vecr_3-\vecr'_3-\vecs)} \Pk(k_3). 
\end{align}

The integral over $\vecs$ can be performed directly, yielding a Dirac delta function that imposes the constraint $\veck_3+\vecq_1+\vecq_2+\vecq_3=0$. This delta function may then be used to eliminate one of the four wave vectors, for example $\veck_3$. At this stage, the brute-force computation involves 4 independent three-dimensional momentum integrals (three \(\vecq\) integrals and one \(\veck\) integral). Discretizing each dimension with \(N\) sampling points implies a total computational cost scaling as:
\begin{equation}
{\rm Cost}_{\rm BF} \sim N^{12}, \nonumber
\end{equation}
which is entirely infeasible for any realistic choice of \(N\). In contrast, our analytical reduction exploits translational invariance, angular-momentum decompositions, and the separability of radial and angular degrees of freedom to rewrite the same covariance contribution in terms of six nested one-dimensional radial integrals. In this case, the computational cost scales instead as:
\begin{equation}
{\rm Cost}_{\rm Reduced} \sim N^{6}, \nonumber
\end{equation}
representing a dramatic reduction in computational complexity. This reduction makes the numerical evaluation of the next-to-leading order 4PCF covariance practical, and is essential for any application to modern large-scale structure data.

\section{Convergence of the Projected Covariance}\label{sec:Convergence_Test}

In order to assess the convergence of the covariance contributions, we project the full covariance onto the isotropic basis by multiplying the results presented in $\S$\ref{sec:Resulting_Cov} by $\mathcal{P}^*_{\Omega}(\hatr_0,\hatr_1,\hatr_2,\hatr_3)$ and $\mathcal{P}^*_{\Omega'}(\hatr'_0,\hatr'_1,\hatr'_2,\hatr'_3)$, and integrating over all angular variables. This removes all angular dependence and allows us to study convergence for specific \emph{external} angular momentum configurations. Projecting the covariance onto this basis yields a scalar quantity for each choice of \emph{external} angular momenta
\begin{align}
\mathrm{Cov}_{\Lambda,\Lambda'; \;(T)}(R,R'), \nonumber
\end{align}
where $\Lambda$ and $\Lambda'$ are \emph{external} angular momenta, $R \equiv \{r_0,r_1,r_2,r_3\}$ and $R' \equiv \{r'_0,r'_1,r'_2,r'_3\}$, and $T \in \{\mathrm{S\text{-}S},\mathrm{P\text{-}S}, \mathrm{P\text{-}P}\}$ labels the covariance contribution.

To study convergence, we introduce an \emph{internal} angular momentum cutoff $L_{\rm cutoff}$, restricting all \emph{internal} angular momenta appearing in the covariance expressions to satisfy $\{L_{\rm int}\} \leq L_{\rm Cutoff}$. The projected covariance truncated at this cutoff is then defined as
\begin{align}
\mathrm{Cov}^{(\leq L_{\rm cutoff})}_{\Lambda,\Lambda'; \;(T)}(R,R')
=
\sum_{\{L_{\rm int}\} \leq L_{\rm cutoff}}
\mathcal{A}_{\{L\},{T}} \;
\mathcal{R}_{\{L\},{T}}(R,R'),
\end{align}
where $\mathcal{A}_{\{L\},{T}}$ collects all angular prefactors and $\mathcal{R}_{\{L\},{T}}$ denotes the corresponding radial integrals. Then, we quantify convergence by comparing the truncated result to a reference calculation performed at a sufficiently large cutoff $L_{\rm cutoff}=10$:
\begin{align}
R(L_{\rm cutoff})\big|_{\Lambda,\Lambda';\,R,R'}
\equiv
\frac{
\mathrm{Cov}^{(\leq L_{\rm cutoff})}_{\Lambda,\Lambda'}(R,R')
}{
\mathrm{Cov}^{(\leq 10)}_{\Lambda,\Lambda'}(R,R')
}.
\label{eq:R_Lcutoff_def}
\end{align}
A value $R(L_{\rm cutoff}) \to 1$ indicates convergence of the internal angular momentum sums.

In this work, we focus on diagonal radial configurations with
$r_1 = r_2 = r_3 = r'_1 = r'_2 = r'_3 = r \in \{50,100\}\,[\mathrm{Mpc}/h]$, and fix $r_0 = r'_0 = 0$. That is, we evaluate the covariance separately for $r = 50\,[\mathrm{Mpc}/h]$ and $r = 100\,[\mathrm{Mpc}/h]$. We remind the reader that $\vecr_0$ and $\vecr'_0$ were introduced to symmetrize the results presented in $\S$\ref{sec:Resulting_Cov}; consequently, they are set to zero when evaluating the covariance. The external angular momenta considered here are also motivated by existing 4PCF analyses, which have thus far focused on relatively low external angular momenta, typically up to $\ell_{\rm max}=4$, in both parity-even and parity-odd measurements. Thus, the configurations studied below are representative of the low-angular-momentum regime most relevant to current 4PCF applications.

\begin{figure}[h!]
    \centering
    \includegraphics[width=0.8 \linewidth]{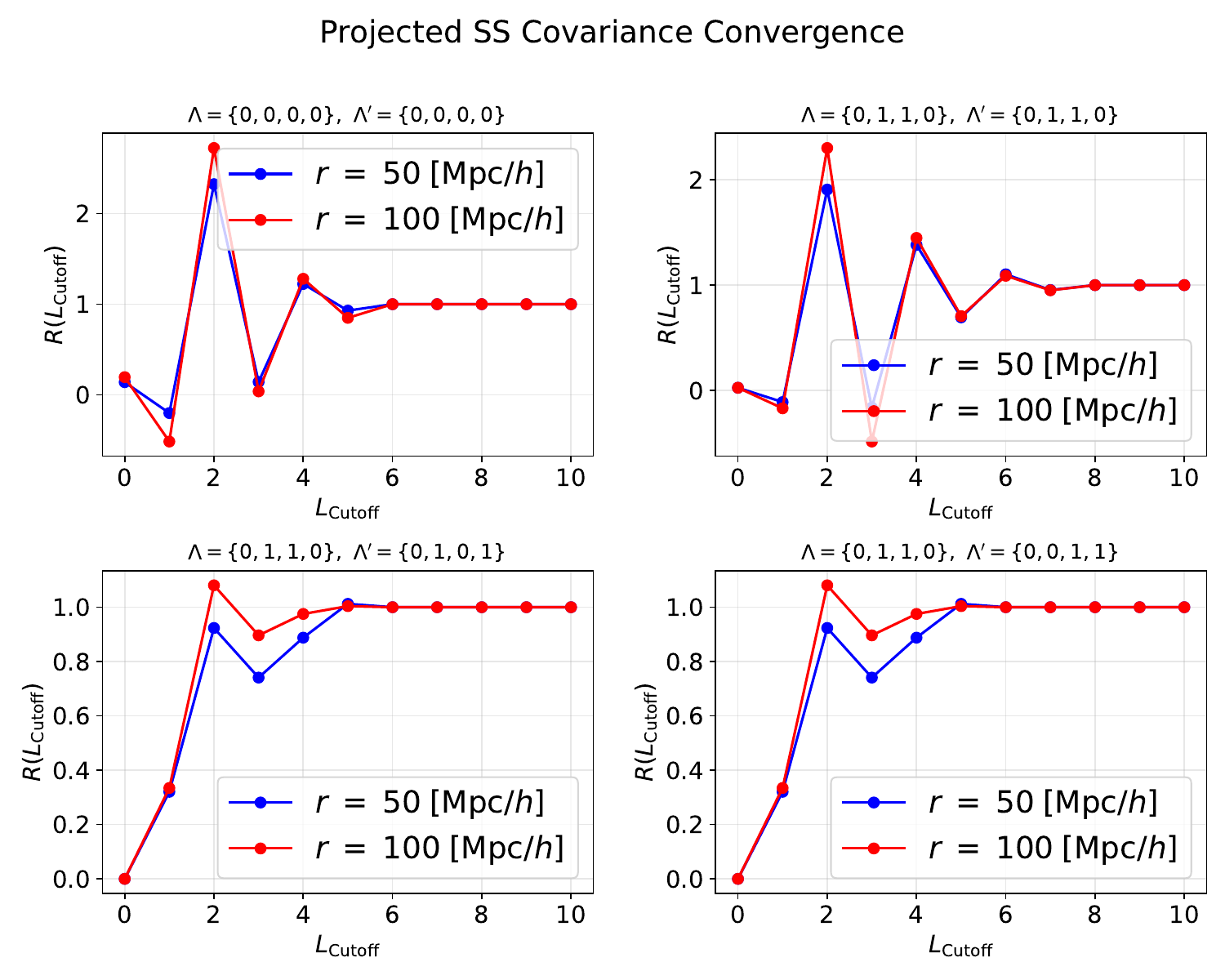}
    \caption{Convergence of the projected S--S covariance as a function of the internal angular momentum cutoff $L_{\rm Cutoff}$ for several fixed external configurations; $\Lambda = \{0, L_1,L_1,L_{33}\}\;{\rm and}\;\Lambda'= \{0, L'_{q21},L'_{q32},L'_{33}\}$. Each panel corresponds to a different choice of external angular momenta, while the lines show the ratio $R(L_{\rm cutoff})$ defined in Eq.~(\ref{eq:R_Lcutoff_def}) for the diagonal radial configurations $r_i = r'_i \in \{50,100\}\,[\mathrm{Mpc}/h]$ and $r_0 = r'_0 = 0$. The ratio approaches unity as $L_{\rm cutoff}$ increases, indicating convergence of the internal angular momentum sums. In all cases shown, the covariance is well converged by $L_{\rm cutoff} =6$.}
    \label{fig:Covaraince_Convergence}
\end{figure}

As an example, Figure~\ref{fig:Covaraince_Convergence} shows the convergence of the projected S--S covariance for several representative external configurations of $\Lambda = \{0, L_1, L_1, L_{33}\}$ and $\Lambda' = \{0, L'_{q21}, L'_{q32}, L'_{33}\}$. Each panel corresponds to a fixed choice of external angular momenta, while the horizontal axis denotes the internal cutoff $L_{\rm cutoff}$. As shown in the figure, the ratio $R(L_{\rm cutoff})$ rapidly approaches unity as the cutoff increases, indicating that only a small number of internal angular momenta
contribute significantly to the covariance. In particular, convergence is
typically achieved by $L_{\rm cutoff} = 6$, demonstrating that the covariance is well converged by $L_{\rm cutoff}=10$ for all configurations considered.

\end{document}